\documentclass[12pt,preprint]{aastex}

\slugcomment{To appear in ApJ}

\shorttitle{2-D Radiative Transfer in Protostars}
\shortauthors{Whitney, Wood, \& Bjorkman}

\begin{document}


\title{
2-D Radiative Transfer in Protostellar Envelopes:  I.  Effects
of Geometry on Class I Sources
}


\author{Barbara A. Whitney\altaffilmark{1}, 
Kenneth Wood\altaffilmark{2}, J. E. Bjorkman\altaffilmark{3},
\& Michael J. Wolff\altaffilmark{1}} 

\altaffiltext{1}{Space Science Institute, 3100 Marine Street, Suite A353, 
Boulder, CO~80303; bwhitney@colorado.edu}

\altaffiltext{2}{School of Physics \& Astronomy, University of St Andrews, 
North Haugh, St Andrews, Fife, KY16 9AD, Scotland the Brave; 
kw25@st-andrews.ac.uk}

\altaffiltext{3}{Ritter Observatory, MS 113, Department of Physics \& Astronomy, 
University of Toledo, Toledo, OH 43606-3390; jon@physics.utoledo.edu}

\begin{abstract}

We present 2-D radiation transfer models of Class I Protostars and show
the effect of including more realistic geometries
on the resulting spectral energy distributions and
images.  We begin with a rotationally flattened infalling envelope
as our comparison model, and add a flared disk and bipolar cavity.
The disk affects the spectral energy distribution most strongly at edge-on 
inclinations, causing
a broad dip at about 10 $\mu$m (independent of the silicate feature)
due to high extinction and low scattering albedo in this wavelength region.
The bipolar cavities allow more direct stellar+disk radiation to
emerge into polar directions, and more scattering radiation to
emerge into all directions.
The wavelength-integrated flux, often interpreted as luminosity,
varies with viewing angle, with pole-on viewing angles
seeing 2-4 times as much flux as edge-on, depending
on geometry.   Thus, observational estimates of luminosity should take into
account the inclination of a source.
The envelopes with cavities are significantly bluer in near-IR
and mid-IR color-color 
plots than those without cavities.  Using 1-D models to interpret
Class I sources with bipolar cavities would lead to an underestimate
of envelope mass and an overestimate of the implied evolutionary state.
We compute images at near-, mid-, and far-IR wavelengths.
We find that the mid-IR colors and images
are sensitive to scattering albedo, and that the flared disk shadows the midplane on 
large size scales at all wavelengths plotted.
Finally, our models produce polarization spectra which can be used to diagnose dust
properties, such as albedo variations due to grain growth.  
Our results of polarization 
across the 3.1 $\mu$m
ice feature agree well with observations for ice mantles covering 5\% of
the radius of the grains..

\end{abstract}


\keywords{radiative transfer---stars: formation---stars: imaging---
stars: pre-main sequence---circumstellar matter---dust, extinction---
polarization}


\section{Introduction}

Lada and collaborators
first classified pre-main sequence stars into an evolutionary sequence,
Class I through III,
by the shapes of their spectra energy distributions (SEDs)
(Lada 1987;  Myers et al. 1987; Adams, Lada, \& Shu 1987).
Class I sources, the subject of this paper, are thought to be T Tauri 
stars surrounded by infalling envelopes.
More recently, an earlier stage was proposed, the Class 0 stage
to describe more deeply embedded sources 
(Andr\'e, Ward-Thompson, \& Barsony 1993).
This called into question whether the Class I envelopes
are merely
remnant or are still collapsing, though detailed radiative
transfer models of several Class I sources
require substantial envelope mass (typically 10-20 times greater than a
Class II disk), and suggest that the envelopes are still infalling, 
if at a less
embedded stage than the Class 0 stage (Adams et al. 1987; 
Kenyon et al. 1993a,b; 
Whitney, Kenyon, \& G\'omez 1997; Stark et al. 2002).

Adams et al. (1987)  successfully interpreted the SEDs of Class I sources
with 1-D radiative transfer models of spherically-averaged 
rotationally-flattened infalling envelopes.
Kenyon et al. (1993a) applied ``1.5-D'' radiative transfer
calculations to model the SEDs of the Taurus Class I sources,
using the spherically-averaged density for calculating the temperature.
The average model infall rates and centrifugal radii were
consistent with 
observed sound speeds and disk sizes.  
Efstathiou \& Rowan-Robinson (1991) also performed accurate 2-D radiative 
transfer solutions using
the rotationally-flattened infalling envelope.
All of these SED models of Class I sources
tend to have similar features, steeply rising at 
about 1 $\mu$m, 
then gradually rising to 100 $\mu$m, and falling off to 1000 $\mu$m
(see Figure 9, top left panel for a similar model).
More recently, Nakazato, Nakamoto, \& Umemura (2003) calculated
2-D models of Class I objects which included a disk and a piece-wise spherically
symmetric envelope divided into two latitude sectors, each having a radial
power law independent of polar angle.  To create a bipolar cavity,
the density was lowered in the polar sector.

Most of the previous 1-D and even the 2-D SED models of Class I sources
underestimate the optical/near-IR 
flux when compared to observations.
Scattered light 
models (Kenyon et al. 1993b) showed that
bipolar cavities are required
to allow an escape path for the short-wavelength radiation, and
to fit the extended images and polarization maps of these sources 
(Whitney et al. 1997, Lucas \& Roche 1997, 1998).  
G\'omez, Whitney, \& Kenyon (1997) found [S II] emission from jets
and Herbig-Haro objects in nearly all the Taurus Class I sources,
providing a mechanism for producing the bipolar cavities.  In addition,
most of the Class I sources are known to have low-velocity molecular
outflows (Moriarty-Schieven et al. 1992, 1995a,b; Chandler et al. 1996,
Tamura et al. 1996, Hogerheijde et al. 1998).  High spatial-resolution
images (Padgett et al. 1999) and scattered-light models
(Stark et al. 2002)
find that vertically extended disks, an expected consequence of rotational 
infall, are required in addition to the infalling envelopes
to fit the images several Class I sources.  The hydrodynamic collapse
models of Bate (1998) also produce substantial disks in Class I sources.

Since it is clear from these studies that
Class I sources have disks and cavities in addition to infalling envelopes,
we wish to determine if these more complicated geometries give different
SEDs than the previous models show.
This paper computes the exact solutions for 2-D radiative transfer
in Class I sources including the realistic geometrical effects of
a rotationally flattened envelope,
a flared accretion disk, and bipolar cavity.  We find that both the disks
and cavities have substantial effects on the resulting SEDs, especially
in the mid-IR.
These variations can be tested with new ground-based and upcoming
space-based infrared observations.
In addition to SEDs, we compute images and polarization.
This paper details how the geometry affects the SED for a single
evolutionary state, the Class I source.
In a followup paper we will explore these effects in an evolutionary
sequence.  
The next section describes the models, \S3 shows the results of four
different geometries, and \S4 concludes with a discussion of the results.

\section{Radiative Transfer Models}

\subsection{Ingredients of the Model: Disk, Envelope, \& Cavity Structure}

We wish to use as realistic a geometry as possible to describe
the Class I protostar.  Since disks appear to be ubiquitous in 
star forming regions (Cohen, Emerson, \& Beichman 1989; Strom et al. 1989;
Beckwith et al. 1990, Dutrey, Guilloteau, \& Simon 1994;
Koerner \& Sargent 1995; Saito et al. 1995; Burrows et al. 1996),
and therefore the infalling envelope likely has angular momentum,
we choose for the envelope
geometry the infall solution that includes rotation
(Ulrich, 1976, Cassen \& Moosman 1981, Terebey, Shu, \& Cassen 1984).  
On larger scales, magnetic fields (Galli \& Shu 1993a,b, Li 1998, Ciolek \&
K\"onigl 1998) and initial motions prior to dynamical collapse
(Foster \& Chevalier 1993) may affect
the structure but within the infalling region, the effects of rotation
likely dominate over magnetic effects.  This region contributes most
to the SED and images.
The density structure for the rotationally-flattened infalling envelope
within the infalling radius is given by:

\begin{equation}
\rho = \frac{\dot M_{\rm env}}{4\pi}\left(\frac{GM_\star} {
R_c^3}\right)^{-1/2} \left(\frac{r}{R_c}\right)^{-3/2}
\left(1+\frac{\mu}{\mu_0}\right)^{-1/2}\left(\frac{\mu}{\mu_0} +
\frac{2\mu_0^2 R_c} {r}\right)^{-1},
\end{equation}

\noindent
where $\dot M_{\rm env}$ is the envelope mass
infall rate, $R_c$ is the centrifugal radius, $\mu = \cos\theta$ and
$\mu_0$ is the cosine polar angle of a streamline of infalling particles as 
$r \rightarrow \infty)$.  The equation for the streamline is given by
\begin{equation} \mu_0^3 +
\mu_0 (r/R_c - 1) - \mu (r/R_c) = 0 \; .
\end{equation}

\noindent
For the disk structure, we use a standard flared accretion density
(Shakura \& Sunyaev 1973, Lynden-Bell \& Pringle 1974; Pringle 1981;
Bjorkman 1997; Hartmann 1998):

\begin{equation}
\rho=\rho_0 
\left [ 1-\sqrt{{R_\star}\over{\varpi}}\right ]
\left( {R_\star \over \varpi} \right)^\alpha\
\exp\left\{{ -{1\over 2} \left[{z\over h(\varpi )}\right]^2  }\right\}\; ,
\end{equation}

\noindent
where $\varpi$ is the radial coordinate in the disk midplane and the
scale height increases with radius,
$h=h_0\left ( {\varpi /{R_\star}} \right )^\beta$.
For the models presented here, we adopt:
$\beta = 1.25$, $\alpha=2.25$, $h_0=0.01R_\star$, giving
a scale height at the disk outer radius of
$h[100{\rm AU}] = 10$AU.
The $\beta$ and $\alpha$ values are consistent with detailed
hydrostatic structure calculations (D'Alessio et al. 1998).   The exact
disk structure in an object with an infalling envelope is likely
to vary from this geometry,
since the disks are accreting from the
infalling envelope.
However, observations and models of Class I sources show morphologies
consistent with vertically extended disks (Padgett et al. 1999,
Stark et al. 2002) so the flared disk seems to be a good starting
point for this investigation.
We include accretion energy, $d\dot{E}_{\rm acc}/dV$,
in the disk using the $\alpha-$disk
prescription (Shakura \& Sunyaev 1973,
Kenyon \& Hartmann 1987, Bjorkman 1997; Hartmann 1998):
\begin{equation}
{d\dot{E}_{\rm acc} \over dV}={{3GM_\star\dot M_{\rm disk}}\over{\sqrt{32 \pi^3} \varpi^3} h(\varpi) }
\left [ 1-\sqrt{{R_\star}\over{\varpi}}\right ]
\exp\left\{{ -{1\over 2} \left[{z\over h(\varpi )}\right]^2  }\right\}\; ,
\end{equation}
where the disk accretion rate, $\dot M_{disk}$ is given by
\begin{equation}
\dot M_{\rm disk} = \sqrt{18 \pi^3}\,\alpha_{\rm disk}\, V_c\, \rho_0\, 
h_0^3/R_\star \; ,
\end{equation}
with the critical velocity $V_c=\sqrt{GM_\star/R_\star}$.
For the disk parameters used in this paper (\S3), Table 1 shows the
resulting disk accretion rate and accretion luminosity.
For simplicity, accretion energy inside the dust destruction
radius, $R_{dust}$, is emitted with the stellar spectrum.  
Thus the accretion luminosity emitted within the dusty disk is
\begin{equation}
L_{\rm acc}= {GM_\star\dot M \over 2R_{\rm dust}} \left[ 3 - 2 \sqrt{R_\star/R_{\rm dust}} \right].
\end{equation}

Since near-IR images of Class I sources require bipolar cavities
to fit the observations (Kenyon et al. 1993b, Whitney et al. 1997),
we include them in these models.
We choose two shapes to
describe the cavities since they both appear to be evident in
observations (Padgett et al. 1999, Stark et al. 2002; Reipurth et al. 2000).  
These are a streamline, which
is conical on large scales, and a curved
cavity.  The streamline cavity shape might occur if precessing jets carve
out a conical shape while infalling material outside of the cavity 
continues to fall in along streamlines.  The cause of a curved shape is less
certain.
The streamline
cavity shape is defined by equation (2); at each $(\mu,r)$, $\mu_0$ is
calculated, and if it is greater than the chosen cavity angle, the
density is set to the cavity density.
Whitney \& Hartmann (1993) show a picture of this cavity shape in their 
Figure 1.
The curved cavity follows $z = a \varpi^b$, 
where $\varpi=\sqrt{x^2 + y^2}$.  We include a small amount of 
dust in the cavity with constant density, 
$n_{H2}= 2 \times 10^4 $cm$^{-3}$,
as expected for a cylindrical outflow.  
The density in the cavity may vary from 
$n_{H2} \sim 10^3-10^6 $cm$^{-3}$
in the
high- and low-velocity flow regions (Bachiller \& Tafalla 1999; Reipurth 
\& Raga 1999).
A more realistic cavity density distribution including such large
variations 
could have a substantial effect on the radiative transfer, and will be
considered in future work.
The total mass in the cavity, about $6 \times 10^{-3} M_\sun$, 
is similar to observed outflow masses
in Taurus Class I sources (Moriary-Schieven et al. 1995b,
Chandler et al. 1996, Hogerheijde et al. 1998).

\subsection{Radiation Transfer Method}

To calculate the radiative transfer,
we incorporated the Monte Carlo 
radiative equilibrium routines developed by Bjorkman
\& Wood (2001) into a 3-D spherical-polar grid code (Whitney \& Wolff 2001).
Bjorkman \& Wood tested their codes
by comparing to a set of benchmark calculations developed by
Ivezi\`c et al. (1997) for spherically symmetric codes.
They then applied their method to 2-D ellipsoidal envelopes and
T Tauri disks (Wood et al. 2002a,b).  These codes use a numerical
optical depth integrator that operates on an analytically prescribed
density.  The numerical integrator has difficulty handling sharp
density gradients.
In order to include arbitrary density distributions, such as bipolar
cavities or 3-D clumpy envelopes,
we have developed a code that uses a 3-D spherical-polar
grid (Whitney \& Wolff 2001).  The density is constant within a grid
cell, allowing for straightforward integration of optical depth through
the grid.  
Since the density in each cell is computed only once at the 
beginning of the processing, we can incorporate very complicated analytic
density formulations, or use tabular
densities derived from hydrodynamical simulations.  
For the protostellar envelope case, the grid has variable spacing
in $r$ and $\theta$ to sample the vastly different density variations and
size scales from
the inner disk region to the outer envelope. 
The radial
spacing is logarithmic in the inner region of the disk, to resolve the inner
edge of the disk in optical depth, 
and then
follows a power-law spacing out to the edge of the envelope, chosen
to be 5000 AU for the models presented here.

We include accretion in our disk, following Wood et al. (2002a).
We also attach a diffusion solution to the inner 
regions (Bjorkman et al. 2003).
We tested our grid code by comparing calculations of ellipsoidal
envelopes and T Tauri disks to the 2-D codes of 
Bjorkman \& Wood (2001) and Wood et al. (2002a,b).   The results
were identical.

In these models, the luminosity is produced by the central
star and disk accretion; it then gets absorbed and reemitted,
or scattered, by the surrounding 
disk and envelope. 
As described in Bjorkman \& Wood (2001), the radiative equilibrium 
algorithm corrects
for the temperature in each grid cell by sampling the new photon frequency
from a differenced emissivity function (the difference between that
computed with the previous erroneous temperature and that computed
from the new corrected temperature).  Energy is conserved and no iteration is
required as the difference spectrum corrects for the previous temperature
errors.  
The scattering phase function is approximated using the Henyey-Greenstein
function, with forward-scattering parameter, $g$, and albedo, $\omega$
computed at each wavelength from our grain model (described below).
Polarization is calculated assuming a Rayleigh-like phase
function for the linear polarization (White 1979).

The Monte Carlo codes produce results simultaneously at all inclinations.
We also incorporate the ``peeling-off'' algorithm
(Yusef-Zadeh, Morris, \& White 1984, Wood \& Reynolds 1999)
to produce high S/N images at a specific inclination in a given run.
Images are produced at several wavelengths and convolved with broadband
filter functions for comparison to observations.  For this paper,
we chose for the near-IR images the Hubble Space Telescope NIC2 
F110W (1.1 $\mu$m), F160W (1.6 $\mu$m), and F205W (2.1 $\mu$m) filters.
For the mid-IR images, we chose the SIRTF IRAC filter functions at
3.6, 4.5, 5.8, and 8.0 $\mu$m.  And for the far-IR images, we chose the SIRTF
MIPS filter functions at 24, 70, 160 $\mu$m.  
A given run produces all the images simultaneously as it produces the SEDs
and polarization spectra.
The models take $\sim30$ minutes to produce 
high quality
spectra (that is, high frequency resolution and low noise) 
at all inclinations on a 1 GHz linux PC running g77 FORTRAN.
To produce high quality images takes several
hours.

\subsection{Dust Model}

The dominant opacity source in the radiative transfer models is dust grains.
Our dust model is based on that derived by Kim, Martin, and Hendry (1994)
for the diffuse ISM (using a code kindly supplied by Peter Martin
and Paul Hendry).
To approximate dust properties in nearby star formation regions,
we produce a size distribution of grains that fits an extinction curve
generated by the Cardelli, Clayton, \& Mathis (1989) prescription, with 
$R_V$=4.
This value of $R_V$ is typical of the denser regions in the Taurus
molecular cloud  (Whittet et al. 2001).
The size distribution is not a simple power law, because it is derived
from a Maximum Entropy Method solution.  The size distribution is similar
to that shown in Figure 1a of Kim et al. (1994).
We adopt the dielectric 
functions of astronomical silicate
and graphite from Laor \& Draine (1993).
To illustrate the effects of ices, which cause observable spectral
features, we include a layer of water ice on 
the grains, covering the outer 5\% of radius.  
We varied this value between
0 and 15\% and found 5\% to agree best with polarization observations,
as discussed in \S 3.5.  
The ice dielectric function is taken from Warren (1984). 

This grain model may be appropriate for the envelope properties but
is likely not applicable in the midplane of the disk where grains should
grow to meter-size within a timescale of $10^4 - 10^5$ years 
(Weidenschilling \& Cuzzi 1993; Wetherill \& Stewart 1993;
Beckwith, Henning, \& Nakagawa 2000).
Thus, at very long wavelengths where thermal emission from the disk
dominates, our SEDs may not agree with observations
(see D'Alessio, Calvet, \& Hartmann 2001, Wood et al. 2002a for models of 
disks using large grains).
A plot of the grain properties vs. wavelength
is shown in Figure 1.   The properties for ISM grains are also shown for comparison.  
The major difference is that the albedo of the $R_V$=4 grains is substantially higher
out to about 10 $\mu$m, which will give brighter images at these wavelengths than ISM grains.

\section{Four Variations on a Class I Source Geometry}

To illustrate the effects of geometry on the SED we will show results
for four geometries.
The first is the rotationally-flattened infalling envelope only, illuminated
by the central star.  This is the geometry used by several previous authors 
(though
spherically averaged in some cases).   
For model parameters, we use typical values
derived from previous models of Taurus sources (Kenyon et al. 1993a,b):
infall rate, $\dot{M} = 5 \times 10^{-6} M_\sun/yr$ , centrifugal radius $R_c = 100$ AU,
envelope outer radius $R_{max}=5000$ AU, and envelope inner radius equal
to the dust destruction radius, about 7.5 stellar radii in this case.
The central star radiates as a 4000 K star, using a Kurucz model atmosphere
(Kurucz 1994), with a radius of $R_\star=2.09 R_\sun$, giving a stellar 
luminosity of 1 $L_\sun$.  
An azimuthal cut through the envelope density structure
is shown in the top-left panel of Figure 2.  It has a toroidal
shape in the mid-plane about twice the size of the centrifugal
radius.  The density profile within this region falls roughly
as $\rho \propto r^{-1/2}$ (see equation 1), and beyond this changes
to $\rho \propto r^{-3/2}$.  

For our second geometry, we add the density structure of the flared
disk (equation 3), as shown in the top-right panels of Figure 2b and 2c.
This disk has a mass of 0.01 $M_\sun$ and an outer radius equal to the
envelope centrifugal radius, 100 AU.
For our disk parameters,
the dust destruction radius is $7.5 R_\star$ which we set as our
inner disk radius. For a disk with
$\alpha_{disk} = 0.01$, and the disk parameters in Table 1,
the fraction of luminosity emitted within the disk (i.e., beyond
$7.5 R_\star$)
is $L_{acc} = 0.0078 L_\star$.
The third geometry includes the disk and a streamline cavity with
an opening angle, $\theta_C$, of 25\arcdeg at 5000 AU.  And the fourth geometry
includes the envelope, disk, and a curved cavity with an opening angle
of 20\arcdeg at 5000 AU.  Notice Figure 2b shows that the curved
cavity carves out more of the inner envelope material, even though
its opening angle at 5000 AU is smaller than the streamline 
cavity. 
Table 1 summarizes all the model parameters
and Table 2 shows the variations for the four models.

Figure 3 shows a plot of optical extinction vs. polar angle through 
the envelope to give an idea of the optical depths involved. 
In Model 1, the rotationally-flattened envelope, the extinction ranges
from A$_V = 32$ pole-on to A$_V \sim 225$ at an inclination of
89.999 degrees (the solution for density
has a singularity in the midplane).  In the
models with bipolar cavities (models 3 and 4), the extinction drops to 
A$_V = 1.5$
near pole-on inclinations.
In the models with disks, the extinctions
increase to several million at edge-on inclinations.  Thus, the
star and inner disk radiation will be extincted out to very long
wavelengths for the edge-on viewing angles in these models.
This radiation can escape through scattering at wavelengths
where the albedo is not negligible.  Wood et al. (2002a)
shows the varying contributions of direct, scattered, and thermal
emission for disk models.  

\subsection{Temperatures}

Figure 4 shows temperatures calculated from the radiative
equilibrium solution in the four geometries.
Some interesting features to note are:  1)  The low-density
bipolar cavities (models 3 and 4) are hotter than the surrounding
envelope, since they are exposed to more direct stellar radiation.
4) The midplane is cooler than the envelope or cavity, especially in
the models with a disk (Models 2-4) (Figure 4b, 4c).
Regions in the midplane beyond about 1 AU see less of the
radiation from the star and warm inner regions due to the large optical
depths through the disk (Figure 3), and thus are heated less than
the surrounding envelope.  4) The interior of the
disk in Model 2 is warmer than those in 
Models 3 and 4 (Figure 4c).  This model has no bipolar cavity and 
thus more radiation is trapped
within the envelope to heat the disk.

Figure 5 shows the temperature
behavior in different regions in more detail.  
This plots temperature as a function
of radius for four polar angles, $\theta=0\arcdeg,30\arcdeg,60\arcdeg$, and 90\arcdeg. 
In Model 1, which has no disk or cavity,
the temperature profiles are similar at all
inclinations.   The equatorial profile is slightly steeper
due to the higher optical depths (A$_V=225$, compared to 33 at the poles).
The temperature scales roughly as $r^{-0.5}$ out to about 3 AU, and
then transitions
to $r^{-0.33}$ beyond 200 AU.  
As shown in Kenyon et al. (1993a), the
temperature profile in optically thin regions is $T \propto r^{-a}$, where
$a = {-2/(4+p)}$,
and $p$ is the power law exponent for the grain opacity, i.e., 
$\kappa_\nu \propto \nu^p$.  Beyond a radius of 100 AU, the emitting 
temperatures are below 40 K, so the peak of the blackbody function is
longward of 75 $\mu$m.  At these wavelengths,
the opacity power law exponent is $p=2$, giving an
optically thin temperature profile of $r^{-1/3}$ in agreement with our 
results.  The light grey line plotted with each model is a power law
with $a=0.33$.
In the inner region, we can attempt to compare to the diffusion solution,
though our opacity exponent varies with wavelength and therefore temperature.
The diffusion solution (Kenyon et al. 1993a) gives 
$T \propto r^{-a}$, where $a = (1+n) / (4+p)$, and
$n$ is
the density power law exponent, $\rho \propto r^{-n}$.  At 1 AU, for the
rotational infall solution, $n = 1/2$.  The temperature is 400 K, giving peak
Planck emission at 7.5 $\mu$m.  At this wavelength there is a kink
in the opacity law, where $p$ varies between 0.5 (on average) 
to the long wavelength
and 1.5 to the short wave.  This gives a range of $a=0.42-0.6$ for the 
calculated temperature
profile.  Our value of $a=0.5$ is within this range.  Kenyon et al. (1993a)
find a value of $a\sim0.7$ in similar envelope geometries as our Model 1;
however, the profiles in Efstathiou \& Rowan-Robinson (1991) appear
to be closer to $a\sim0.5$.  These differences could be due in part
to different grain models and envelope parameters.

The temperature profiles for models with disks and cavities have much
more variation with polar angle, as shown in Models 2-4 in Figure 5.
The midplane temperature profile shows a steep
drop in the disk, with $T \propto r^{-1.15}$ out to 1 AU, and flattening
out to 100 AU.  
Then it follows a similar profile as the rest
of the envelope, scaled to a lower temperature.  
The temperature profile in the polar
region in the models with cavities (Models 3-4) is similar to the 
models without cavities.  
It flattens out to the optically thin profile, $a=1/3$,
at a smaller radius, leading to higher overall temperatures in
the cavity.  The 30\arcdeg and 60\arcdeg profiles (dotted and dashed lines) 
diverge from the 0\arcdeg~line where the 
path intersects
the cavity wall, and drops to the lower envelope scale.  This happens
earlier in Model 3 than Model 4 due to the shape of the cavity walls.
At large radii, the models with cavities are hotter along all directions,
even in the midplane.

To better understand the temperature behavior in the disk, in Figure 6
we compare the midplane temperature profile of the models with disks
(Models 2-4) with those of
disks with no envelopes.  The grey line plotted with
each model is a power law, 
$T = 90 \, (r / 1 {\rm AU})^{-0.4} \, {\rm K}$.  
It is clear from these plots that no single power law characterizes
the temperature behavior in either the disk-only or disk+envelope
models.
In all of the models
except Model 2, the temperatures are similar between about 2-10 AU.
In this region, the accretion luminosity is not as important, with
the disk temperature determined mostly from
stellar flux heating the flared surface.
The midplane temperature is higher in Model 2 than any of the others 
due to extra heating from the surrounding envelope (see also
D'Alessio, Calvet, \& Hartmann 1997).
For Models 3 and 4, the cavities are not providing much additional
heating to the disk. As we will show in the next section, the emissivity
in the cavities is actually higher than in the same region of the envelope in Model 2.  
However, this radiation is also able to escape through the cavity
in Models 3 and 4, whereas the envelope radiation in Model 2 is
trapped and heats up the disk.
Beyond about 10 AU, Models 3 and 4 diverge from the disk-only models
due to extra heating from the surrounding envelope.  
In this region,
Model 3 is even higher than Model 2, probably due to heating of the 
cavity walls by stellar radiation (due to their higher curvature, they
intercept more stellar radiation).

Inside 1-2 AU, the temperature profiles
are the same for all the models with disk accretion, with a power
law exponent $a=1.15$.
This is higher than expected for passive accretion ($a=0.75$; Kenyon \&
Hartmann 1987), but
is in agreement 
with the plotted profiles of D'Alessio et al. (1998) between 0.1 and 2 AU.
The temperature is determined by the accretion luminosity
which has a steep falloff with radius (equation 4).
Between 0.073 and 0.16 AU (7.51 to 16.4 $R_\star$), the accretion luminosity
raises the disk temperature above the dust sublimation temperature
of 1600 K.  We set the opacity to 0 in this region since the gas opacity
is much lower than dust.  With surface density of 
1300-1400  ${\rm g} \, {\rm cm}^{-2}$
and 
gas Rosseland mean opacity 
$\kappa_R \sim 10^{-4} \, {\rm cm}^{2} \, {\rm g}^{-1}$
(Alexander, Johnson, \& Rypma 1983), the vertical gas optical depth through the
disk (top and bottom) is $\tau = \Sigma \kappa \sim 0.1$ which is
reasonably optically thin (Lada \& Adams 1992).
For the disk without accretion, the temperature profile
exponent is $a=0.6$ from 0.1-1 AU, then flattens out to $a=0.4$, and even
lower beyond 10 AU.  This is in reasonable agreement with blackbody
emission models that give $a=0.5$ for our flared disk with $\beta=5/4$
(Kenyon \& Hartman 1987).

\subsection{Emissivity and Infrared Images}

Even though Figures 4 and 5 show that temperatures in the bipolar cavities
are higher than in the envelope,
we might expect the overall emitted energy from the cavities to be small due 
to the low densities.  However Figure 7 shows
that the high temperatures lead to large emissivities
in the cavity, despite the lower densities.  
This is especially apparent
at the longer wavelengths, 24 and 160 $\mu$m, shown here.
The emissivity is defined as $j_\nu = \kappa_\nu \rho B_\nu$.  Figure 7
plots azimuthal slices of the emissivity.
It shows how much energy per unit volume is emitted throughout 
the envelope (but
is not a radiative transfer calculation showing how light is propagated
through the envelope).
Even though the models with bipolar cavities have slightly lower mass than
the others, their total emissivities are larger due to the higher temperatures
in the cavities.
Note that the images are plotted to different
size scales for the three wavelengths.  Though the magnitude
of the emissivity is much lower at longer wavelengths,
the brightness profile is much shallower.  

Figure 8 shows images that result from the radiation transfer
calculations.  
These
are shown at an inclination of $i=80$\arcdeg.
From top to bottom, the four model geometries are plotted,
and from left to right, 3-color composities of near-IR (left), mid-IR (middle),
and FIR (right) are plotted.  The images are displayed to mimimum flux
levels 
described in the figure caption,  
assuming a distance to the source of 140 pc.
The near-IR images are similar to observations and previous models
(Padgett et al. 1999, Stark et al. 2002).
The mid-IR images should be detectable by the upcoming Space
Infarared Telescope Facility (SIRTF) in nearby star forming
regions, and the FIR images should be detectable by SIRTF out
to much larger distances.  
As Figure 7 showed, the mid-IR {\it thermal} emission should be rather 
localized near the
central source.  The large extent of the mid-IR images in Figure 6 results 
from {\it scattered} light.
As discussed in Figure 1, the albedo of the $R_V=4$ grains used here is
higher than ISM dust at these wavelengths.  
Since the albedo of the dust is sensitive to grain size, the
mid-IR images should be a useful
diagnostic of grain size.
The FIR images delineate cavity walls and show shadowing
in the disk.  However, these effects will be completely washed out
at SIRTF resolution,
which, at 70 $\mu$m, has a full width at half maximum (FWHM) of 38\arcsec or
2380 AU at a distance of 140 pc.
We convolved the images with Gaussian point spread
functions broadened to the HST and SIRTF resolution for each wavelength.  
We don't show them here because the near-IR
and mid-IR images are essentially unchanged at the resolution shown here,
and the FIR images are washed out and indistinguishable from each other, though
they are elongated with about a 2/1 ratio.

\subsection{Spectral Energy Distributions}

The model SEDs are shown in Figure 9.  
Ten inclinations are plotted, the top curve corresponding
to pole-on and the bottom to edge-on.   
The absorption feature at 3.1 $\mu$m is due to ice opacity, and that
at 10 $\mu$m is due to silicate.
A distinctive feature of the SED plots is the broad
dip at 10 $\mu$m at low inclinations in models that include
the flared disk.  This is due to the disk geometry and the grain
properties:  At shorter wavelengths, even though the opacity through
the disk and envelope is larger (Figure 3), the albedo increases 
allowing light to scatter out at all inclinations.
Note that the near-IR colors are determined completely by scattering
for inclinations $i>60$ ($\cos i < 0.5$), so
the concept of ``de-reddening'' based on the colors is meaningless
since that only applies to direct extincted light, not scattered
(see Stark et al. 2002).
At wavelengths longward of 10 $\mu$m, the opacity decreases enough
to allow thermal emission to propagate through the disk and
envelope.  Near 10 $\mu$m, the low albedo and high opacity combine to
prevent radiation from escaping at high inclinations.

Another striking feature of the SEDs in Figure 9 is that the models
with bipolar cavities (Models 3 and 4)  are much brighter 
at optical/near-IR wavelengths
in better agreement with observations than the models without cavities
(Kenyon et al. 1993a,b).  
Notice in Models 3 and 4 that the pole-on inclinations still
suffer extinction at short wavelengths due to the dust in the cavity.
The shortwave clearly is very sensitive to the amount of dust in the cavity.
Models 3 and 4 also show the 10$\mu$m silicate feature in emission at pole-on
inclinations, due to low optical depth to the source of emission (the inner
disk surface).

It is clear from the plots in Figure 9 that
the more edge-on spectra have lower integrated
fluxes than other inclinations.
In a spherical source, the integral of the measured SED multiplied
by $4\pi d^2$, where $d$ is the distance to the source, is
equal to the luminosity of the source.
However, this is not true for sources with non-spherical envelopes,
where the SED must be integrated over direction as well as frequency,
i.e., 
\begin{equation}
L=2 \pi d^2 \int_{\mu = -1}^{1}{{\int_{\nu=0}^{\infty}}{F_\nu d\nu d\mu}},
\end{equation}
where $\mu = \cos i$ and $i$ is the inclination, also the polar angle.
We note that in our models, this integral is equal
to our input luminosity to within 0.02\% (the difference is
due to performing the integral after coarsely binning the results in angle, 
since flux is fully conserved in our models).
Figure 10 shows the
correction factor that would have to be applied to the integrated
SED at a single inclination to get the true luminosity of the source.  
The correction factor is simply the luminosity divided by the
integrated SED at each inclination $i$,
\begin{equation}
{\rm correction} = 
\frac{ \frac{1}{2} \int_{-1}^{+1} \int_{\nu} F_{\nu} d\nu d\mu }
     { \int_{\nu} F_{\nu}(i) d\nu }.
\end{equation}
As Figure 10 shows, the
observed SEDs can estimate the luminosity of a source only
to within a factor of 2.  While it is not surprising that the
edge-on integrated flux is lower than the true luminosity, it is interesting
to note that the pole-on integrated flux is a factor of 2 larger than
the true luminosity for the sources with bipolar cavities.  This inclination
receives a maximum of direct stellar flux,  scattered flux and reprocessed
emission.  

Nakazato et al. (2003) defined a parameter, $f_L$, as the ratio of
the integrated SED,
$\int{F_\nu d\nu}$ to the peak flux $\nu F_\nu$, and argued that
this can be used as an indicator of inclination.  
We had to modify their definition
slightly because for our models with cavities, the wavelength of the peak
shifts from about 70 $\mu$m to 1.6 $\mu$m at the pole-on inclinations,
causing $f_L$ to decrease for pole-on inclinations.
We can resolve this by defining $f_L(FIR)$ to be the ratio
of the SED to the peak flux in the far infrared (FIR).  
Figure 11 plots $f_L(FIR)$
vs. inclination for our set of models.  
Our models also show a correlation
with inclination, but the behavior is different from Nakazato et al.'s
values.  This is due to the differences in our model geometries.  
Nakazato et al. include an envelope, disk and cavity in their models
but their envelope is spherically symmetric up to the cavity opening
angle.  Thus, their $f_L$ is relatively flat up to the cavity opening angle
and then flat within the cavity at a high value.
In contrast, our models using a rotationally flattened envelope show
a more gradual increase in $f_L(FIR)$ as the inclination becomes more pole-on
(increasing $\cos i$), due to varying envelope extinction with polar angle,
a different cavity shape, and a different assumption for the dust
content in the cavity.  Thus, the behavior of $f_L$ with inclination
depends strongly on model assumptions.
Nakazato et al. claim this parameter $f_L$ can determine
inclination to $\pm5$\arcdeg, and use as an example IRAS 04365+2535
for which they find an inclination of 20\arcdeg.  However, the high
near-IR polarization 
($P_K \sim 10-15\%$; Whitney et al. 1997, Lucas \& Roche 1998),
and molecular outflow observations (Chandler et al. 1996) suggest
inclinations of $i \sim 40\arcdeg-70\arcdeg$.  We suggest that more detailed modeling,
including as much observational information as possible,
is required to ``calibrate'' the models before applying this factor
based on SED information alone.

\subsection{Color-Color Plots}

Figure 12 shows color-color plots for our models.   
The black crosses are main sequence/giant branch stars, and the dotted
lines extending from them show the reddening lines.  The region enclosed
by these lines shows where the MS/GB falls on these plots.
The model points are plotted with different shades of grey representing
inclination, with light grey representing pole-on, and black edge-on.
The models without cavities are shown by the filled circles (Model 1)
and asterisks (Model 2).  The models with cavities are shown by
open triangles (Model 3) and open diamonds (Model 4).  
The models with cavities are substantially bluer in all the plots.
In a source observed to have a bipolar cavity (as indicated
by images resembling Figure 8, outflows, and high polarization),
it is clear from Figure 12 that 2-D cavity models should be used to interpret
their colors.  Using 1-D models, or models without bipolar cavities, would lead
to assignment of a later evolutionary
state to the source than is appropriate.
Another interesting feature in Figure 12 is that some of the edge-on sources
are the bluest.
This is
 because the disk blocks all the direct light, leaving only scattered
light which is blue, especially in the MIR where the albedo increases
steeply towards shorter wavelengths.  

The bottom left panel of Figure 12 shows that the Class I source colors
are well-separated from reddened main sequence and red giant stars for
this particular color choice in the mid-IR and our grain model. 
The bottom right panel, plotting [3.6]-[5.8] vs [8.0]-[24.0] shows
that Class I sources are very red for this color choice.  The main sequence
stars have to be reddened by at least A$_V \gtrsim 100$ to overlap
with these sources.  Thus, in nearby star forming clouds, where foreground
main sequence stars have almost no reddening, the Class I sources will
likely stand out.  Our next paper will explore in more detail 
how well different evolutionary states are separated
in color-color space.

\subsection{Polarization}

An additional tool to distinguish the various geometries as well as
grain properties
is the polarization spectrum, shown in Figure 13.
Plotted is $Q=P\cos(\chi)$, where $\chi$ is the orientation of the
polarization vector.  Because these models are axisymmetric,
and the grains are spherical (no dichroism),
the integrated or net
polarization is either parallel or perpendicular to the rotation
axis.  If $Q/I$ is less than 0, the net polarization is perpendicular
to the rotation axis.  This is the case over much of the spectrum
because most of the scattered light emerges from the bipolar cavity
regions.  
In all of the models, the polarization is highest for the edge-on
inclinations and decreases to 0 pole-on.
The polarization is sensitive to the envelope optical depth:
To get high polarization from scattered light
requires optical depths to the unpolarized stellar/disk
radiation of $\tau > 5$ (Whitney et al. 1997).  We see
this transition in the  1-10 $\mu$m region, depending on inclination.
At longer wavelengths, more scattered radiation emerges from
the equatorial region because the optical depths in the polar regions
are too low. 
We should expect to see the polarization reverse sign at
these wavelengths.  However,
the grain albedo is decreasing rapidly 
in this region so the polarization vanishes before reversing
sign (though there is a hint in Model 4 of a polarization reversal).
In a less embedded source we would expect the polarization
to reverse sign at a shorter wavelength so the polarization reversal
would occur before the albedo drops to 0.  
We will explore this in more detail in future papers.
In disks, where the grains are larger, the albedo should remain
higher towards longer wavelengths (Wood et al. 2002a).
Thus, polarization spectra should be a useful diagnostic of grain growth
as well as evolutionary state.

Figure 13 also shows the behavior of polarization across the
ice band feature at 3 $\mu$m.  The polarization increases
across these features because the opacity increases.
With higher opacity, less direct (unpolarized)
light is present to dilute the polarized light.
This behavior agrees with observations across the
3$\mu$m feature (Hough et al. 1996, Kobayashi et al. 1999, Holloway
et al. 2002).

\section{Conclusions}

We have shown that 2-D models of realistic Class I source geometries produce
substantially different SEDs from 1-D and simplified 2-D models.  The inclusion
of a flared disk and bipolar cavity changes the
SEDs and colors substantially.  The effect of the disk is to cause a
broad dip near 10 $\mu$m in nearly edge-on sources ($i > 80\arcdeg$), and the cavity
allows more shortwave radiation to escape at all inclinations.  
Both the disk and cavity introduce
inclination effects much greater than the simple rotational infall 
model gives (Model 1). 
In addition to the spectral shape, the integral of the SED
also varies with inclination.
In 1-D models, this integral is equal to the luminosity of the source.  In 2-D
or 3-D models, the flux must be integrated over direction as well,
as shown in equation (7).  This is impossible to do for an observed
source which is only viewed from one direction.  Figure 10 shows
the correction that would have to be applied to a source with the
same geometry as our models to
get the true luminosity of the source for a given inclination.


Nakazato et al. (2003) claim that the inclination can be determined from
a factor $f_L$ which is the ratio of the integral of the SED to the peak
of $F_\nu d\nu$.  Our models give a different behavior for $f_L$ due
to the different envelope geometry.  We conclude that $f_L$ on its own
is not a robust indicator of inclination.  Perhaps
incorporating many more observational constraints to the models could
provide some calibration to this factor.

We placed our model results on color-color plots (Figure 12)
for easier comparison to
observations, and find that the models with bipolar cavities
are much bluer than those without.  In addition, the models
with cavities show a large spread with inclination.
It is clear from these plots that realistic 2-D models are required
to properly interpret the colors of Class I sources with bipolar cavities.
A 1-D model would require a lower envelope mass to fit the same colors,
which would imply a later evolutionary state for the source than is
appropriate.

The images in Figure 8 and polarization spectra in Figure 13
can be used to understand dust properties in star forming regions.
Our dust model, fit to the extinction curve typical in the Taurus
molecular cloud, has enough scattering albedo in the mid-IR to
produce extended scattered-light images in the mid-IR.  As shown
in Figure 1, this albedo is higher than the ISM value.  These images
are most easily detected at edge-on inclinations where the 
flared disk blocks the
bright direct stellar and inner-disk radiation.  The sensitivity of SIRTF may
be enough to image these sources and confirm the higher albedo due
to larger grains.  The polarization spectra also provide constraints
on both envelope geometry and grain properties.  Our choice
of an ice mantle occupying 5\% of the outer radius of the grains gives
a 3.1 $\mu$m feature in good agreement with observations.  
Further models at different evolutionary states will be needed to
better characterize the polarization behavior in protostars.

Given that bipolar cavities and inclination effects introduce a large
spread in observational parameters for Class I sources, we suggest
that images, outflows, and polarization measurements can aid
in determining cavity geometry and inclination.  
Then 2-D models can
more accurately assign envelope mass and evolutionary states to 
Class I sources.  In our next paper we will explore in more detail
the effects of evolution on the SEDs, colors, images, and polarization
of protostars.


\acknowledgments

We thank Martin Cohen for providing the colors of main sequence/giant branch
stars in Fig. 12, as well as providing zero-point magnitude scales for 
several SIRTF wavebands.
This work was supported 
by the National Science Foundation (AST-9909966, AST-9819928)
and the NASA Astrophysics Theory Program (NAG5-8587).  K. Wood acknowledges
support from the UK PPARC Advanced Fellowship.





\clearpage


\begin{deluxetable}{lll}
\tablenum{1}
\tablewidth{0pt}
\tablecaption{Model Parameters}
\tablehead{
\colhead{Parameter} &
\colhead{Description} & \colhead{Value} }
\startdata
$R_\star$       & Stellar radius                   & 2.09 $R_\sun$ \\
$T_\star$       & Stellar temperature              & 4000 K \\
$L_\star$       & Stellar luminosity               & 1 $L_\sun$ \\
$M_\star$       & Stellar mass	                   & 0.5 $M_\sun$ \\
$M_{\rm disk}$      & Disk mass                        & 0.01 $M_\sun$ \\
$\alpha$        & Disk radial density exponent     & 2.25 \\
$\beta$	        & Disk scale height exponent       & 1.25 \\
$h_0$           & Disk scale height at $R_\star$   & 0.01 \\
$R_{\rm d,min}$     & Disk inner radius                & 7.5 $R_\star$ \\ 
$R_{\rm d,max}$     & Disk outer radius                & 100 AU\\ 
$\alpha_{\rm disk}$ & Disk Viscosity parameter 	   & 0.01 \\
$L_{\rm acc}(R_{\rm d,min})$       & Disk accretion luminosity to $R_{d,min}$       & 0.0078 $L_\star$ \\
$\dot M_{\rm disk}$ & Disk accretion rate  &  $1.4\times10^{-8} M_\sun/$yr \\
$R_{\rm e,min}$     & Envelope inner radius            & 7.5 $R_\star$ \\ 
$R_{\rm e,max}$     & Envelope outer radius            & 5000 AU\\ 
$\dot M_{\rm env}$  &  Envelope infall rate   & $5\times10^{-6}$ $M_\sun/$yr \\
\enddata
\end{deluxetable}

\begin{deluxetable}{llllll}
\tablenum{2}
\tablewidth{0pt}
\tablecaption{Model Variations}
\tablehead{
\colhead{Name} &
\colhead{Disk} & \colhead{Cavity} & \colhead{Cav. shape} &
\colhead{$\theta_C$} & \colhead{Env. Mass}
}
\startdata
Model 1  &  n & n & $\ldots$    & $\ldots$ & $0.189 M_\sun$\\
Model 2  &  y & n & $\ldots$    & $\ldots$ & $0.189 M_\sun$\\
Model 3  &  y & y & Streamline  & 25\arcdeg & $0.177 M_\sun$\\
Model 4  &  y & y & Curved      & 20\arcdeg & $0.176 M_\sun$\\
\enddata
\end{deluxetable}


\clearpage



\begin{figure}
\figurenum{1}
\epsscale{1.0}
\plotone{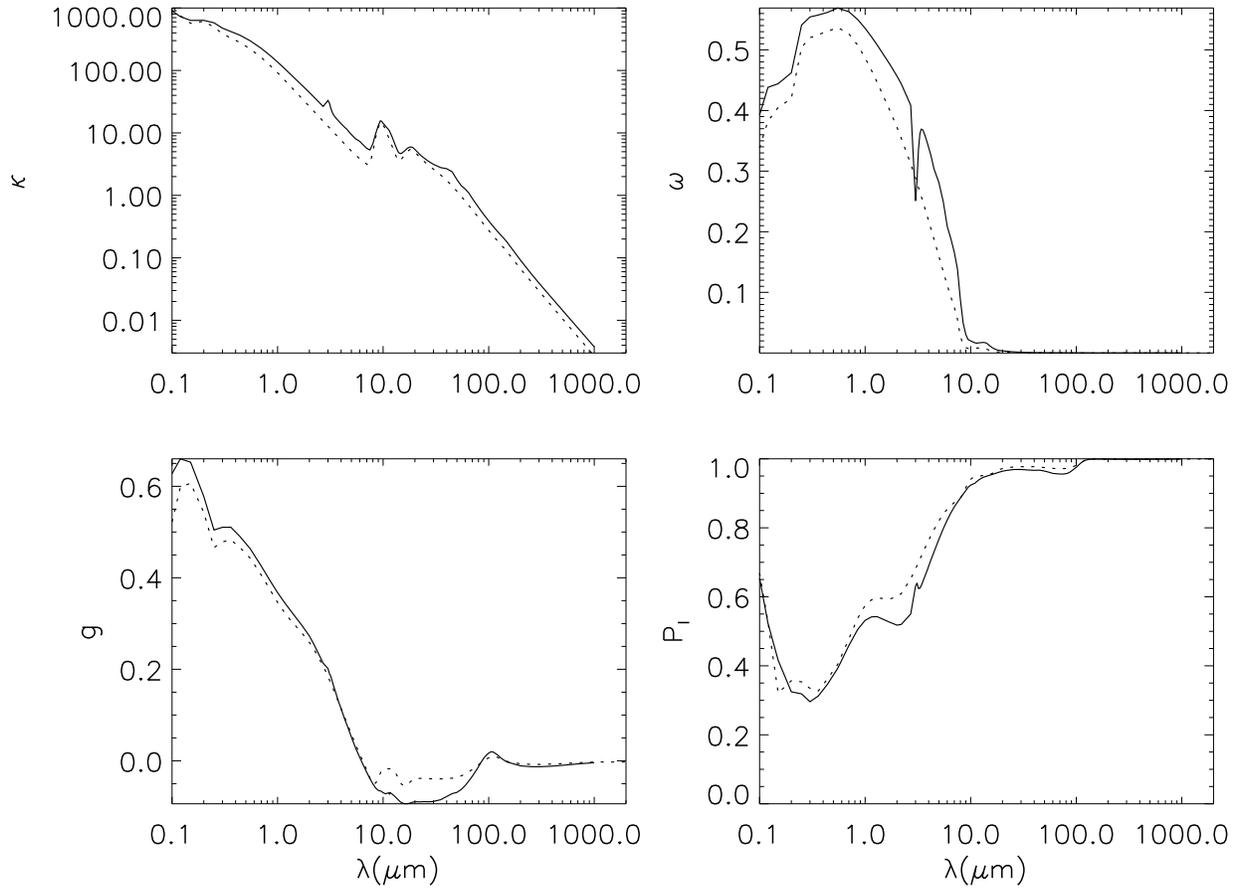}
\caption{Behavior of the $R_V=4$ grains with wavelength.  For comparison, the dotted lines show behavior of ISM grains ($R_V=3.1$).Top left:
opacity, $\kappa$; top right: albedo, $\omega$; bottom left: average cosine of
scattering angle, $g$; bottom right: maximum polarization, $p_l$.  
}
\label{dust}
\end{figure}
\clearpage

\begin{figure}
\figurenum{2}
\epsscale{1.0}
\plotone{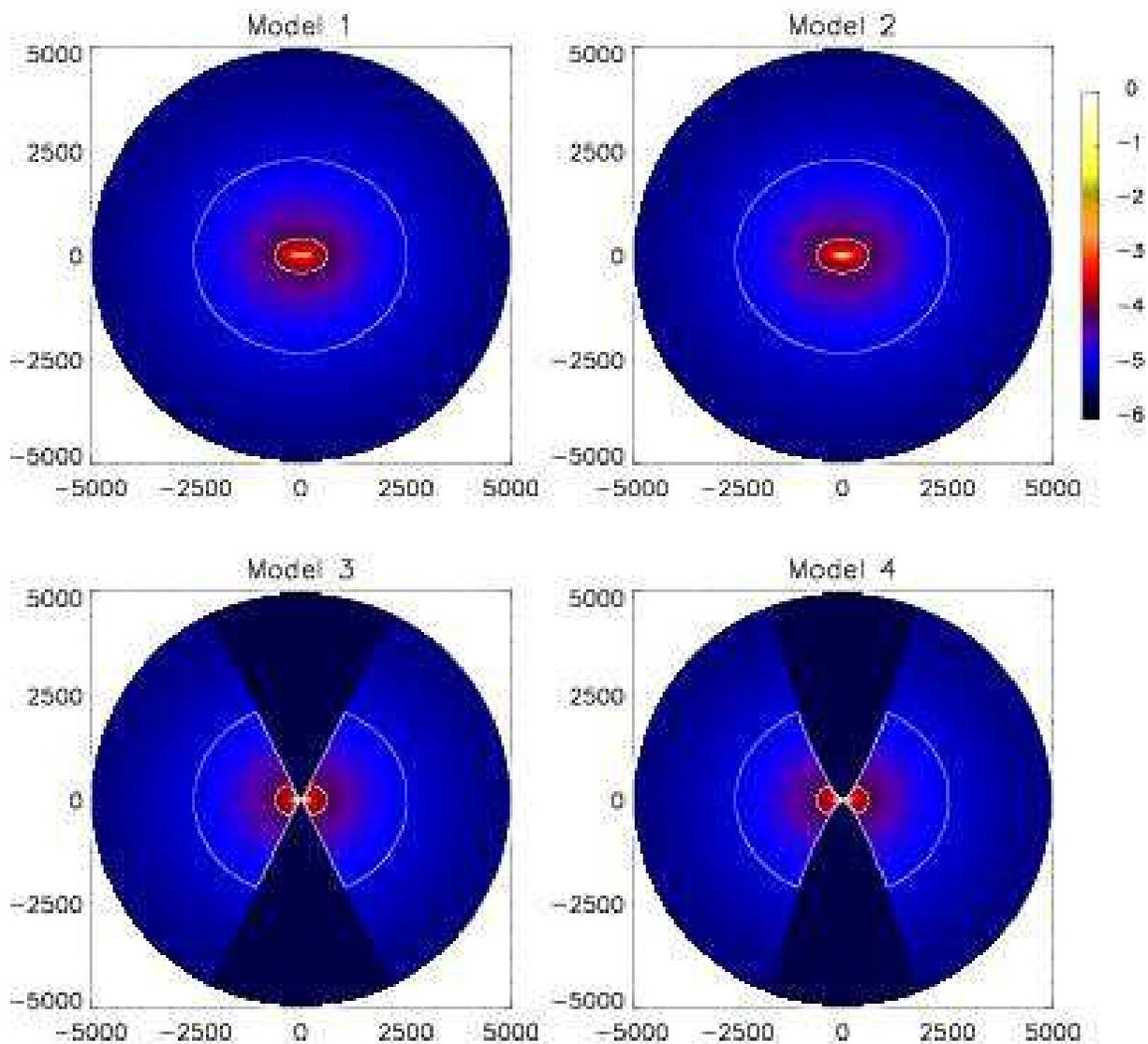}
\caption{Density plots showing the four geometries.  Model 1:  
rotationally flattened infalling envelope;  Model 2:  
same as Model 1 + flared disk; 
Model 3: same as Model 2 + streamline cavity; Model 4: same as Model 2 +
curved cavity.
The scaling is logarithmic, plotted to six orders of magnitude
from the peak shown.  The contours each decrease by 1
order of magnitude, and match the labels of the color bar.
The plotted regions and peak densities are
(a) 5000 AU, $3 \times 10^{-14} $gm cm$^{-3}$,
(b) 300 AU, $1.5 \times 10^{-12} $gm cm$^{-3}$, and
(c)  20 AU, $8 \times 10^{-12} $gm cm$^{-3}$.
}
\label{dens5000}
\end{figure}
\clearpage

\begin{figure}
\figurenum{2b}
\epsscale{1.0}
\plotone{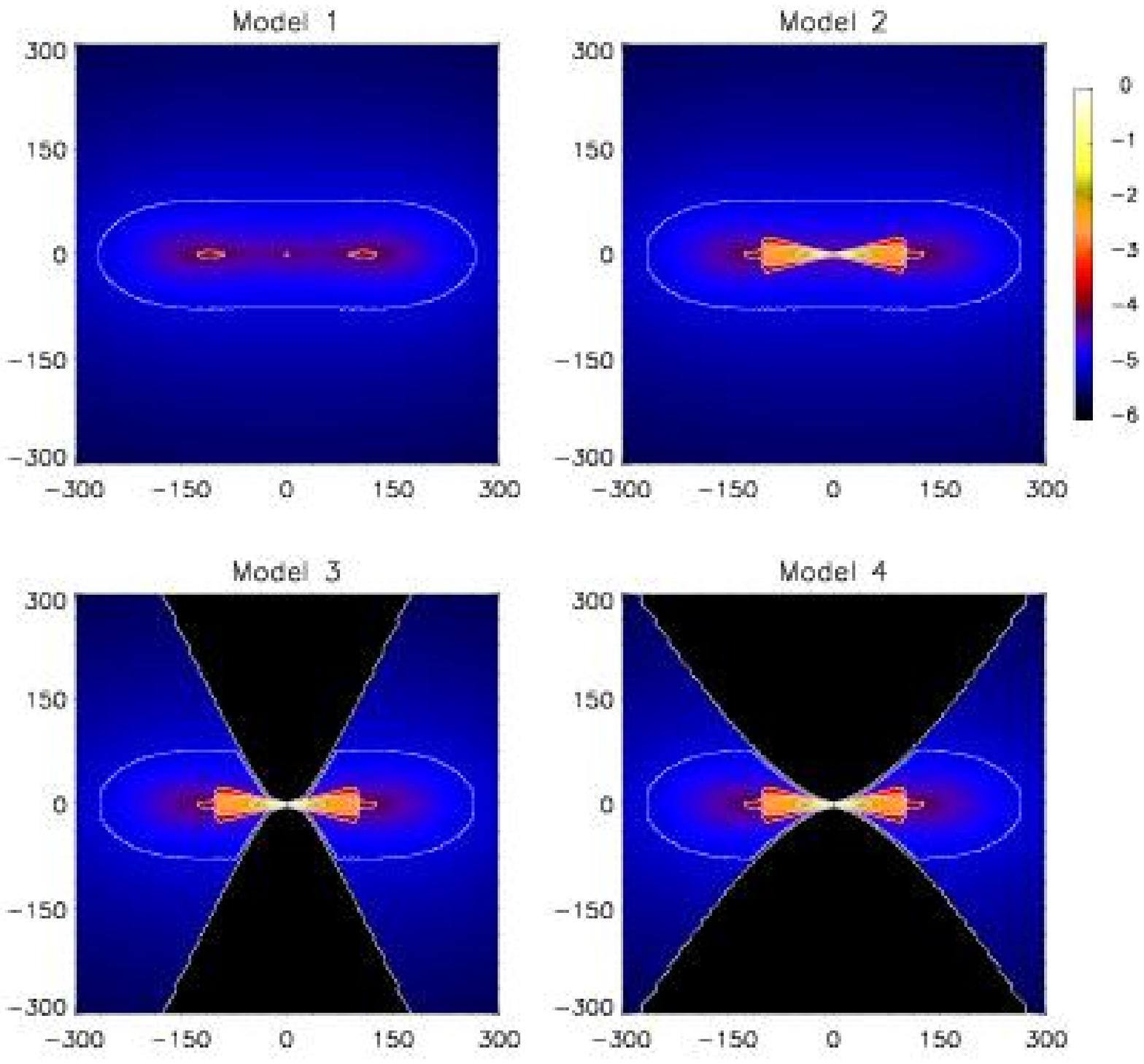}
\label{dens200}
\end{figure}
\clearpage

\begin{figure}
\figurenum{2c}
\epsscale{1.0}
\plotone{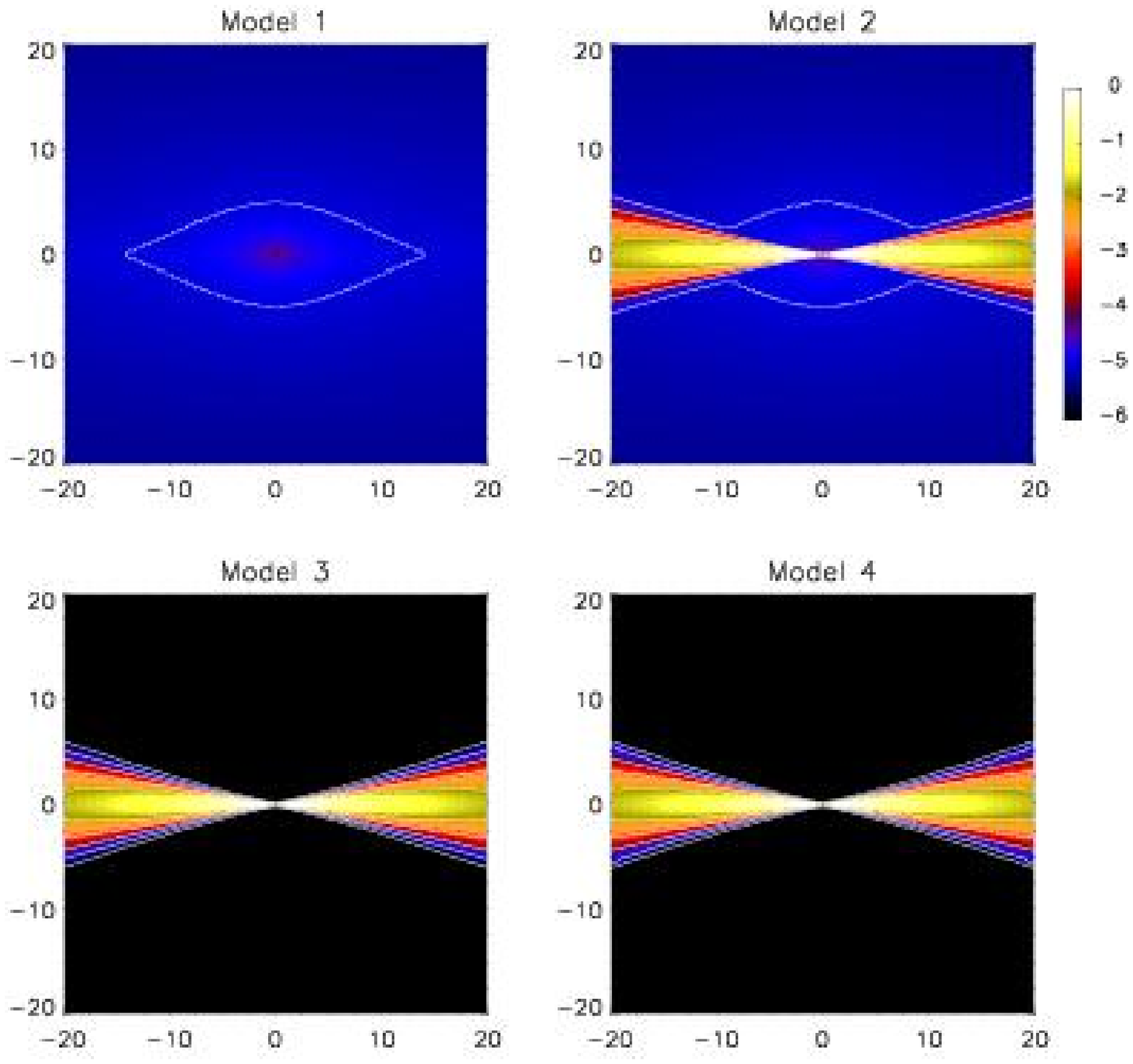}
\label{dens200}
\end{figure}
\clearpage

\begin{figure}
\figurenum{3}
\epsscale{1.0}
\plotone{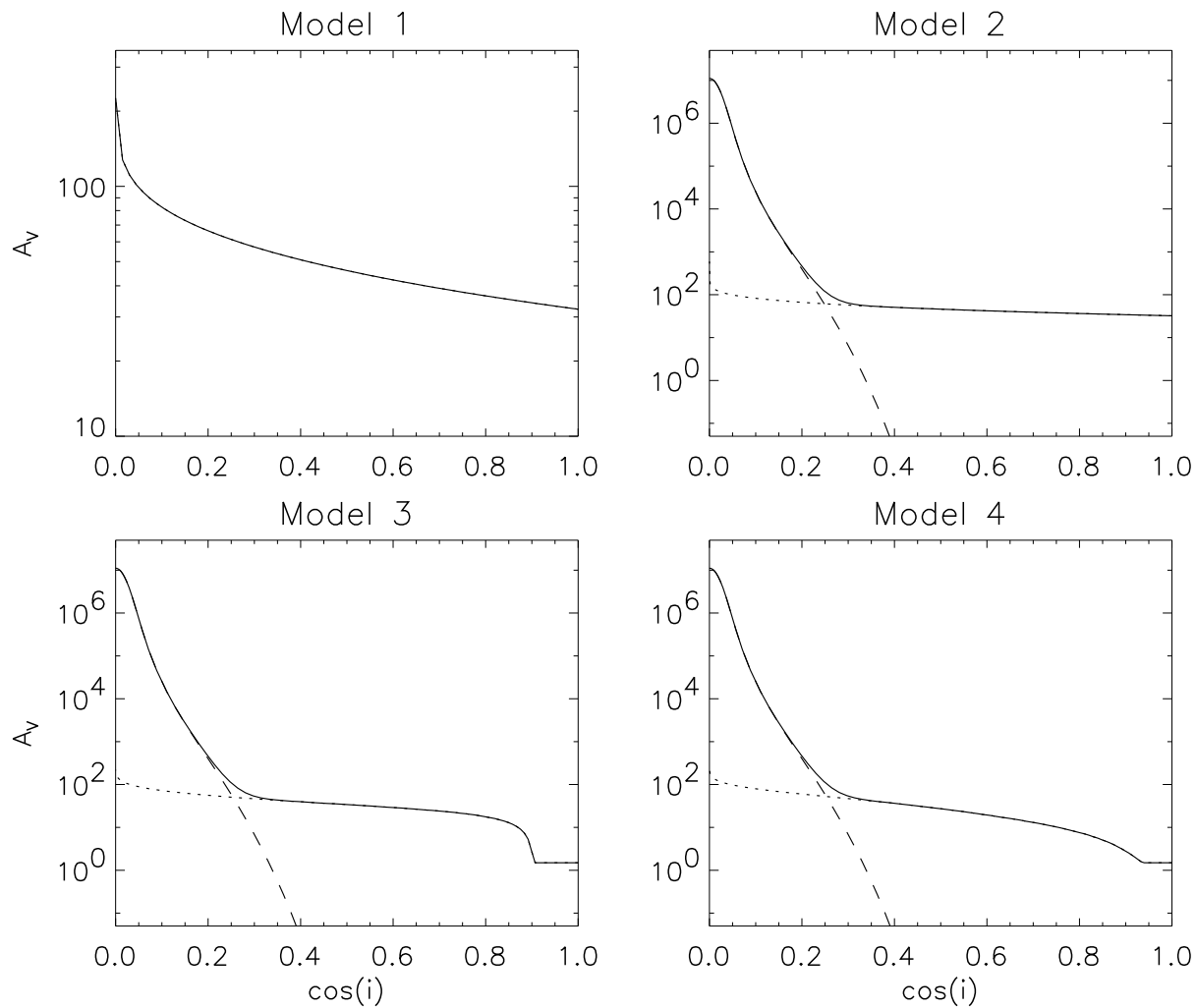}
\caption{Optical extinction, A$_V$, from the central star through the envelope
vs. inclination with respect to the polar axis, $i$.  The dotted line
shows extinction due to the envelope, the dashed line is that of the disk, and
the solid is the sum.  Note the large extinctions at edge-on inclinations
for models with disks (Models 2-4).
}
\label{taui}
\end{figure}
\clearpage

\begin{figure}
\figurenum{4a}
\epsscale{1.0}
\plotone{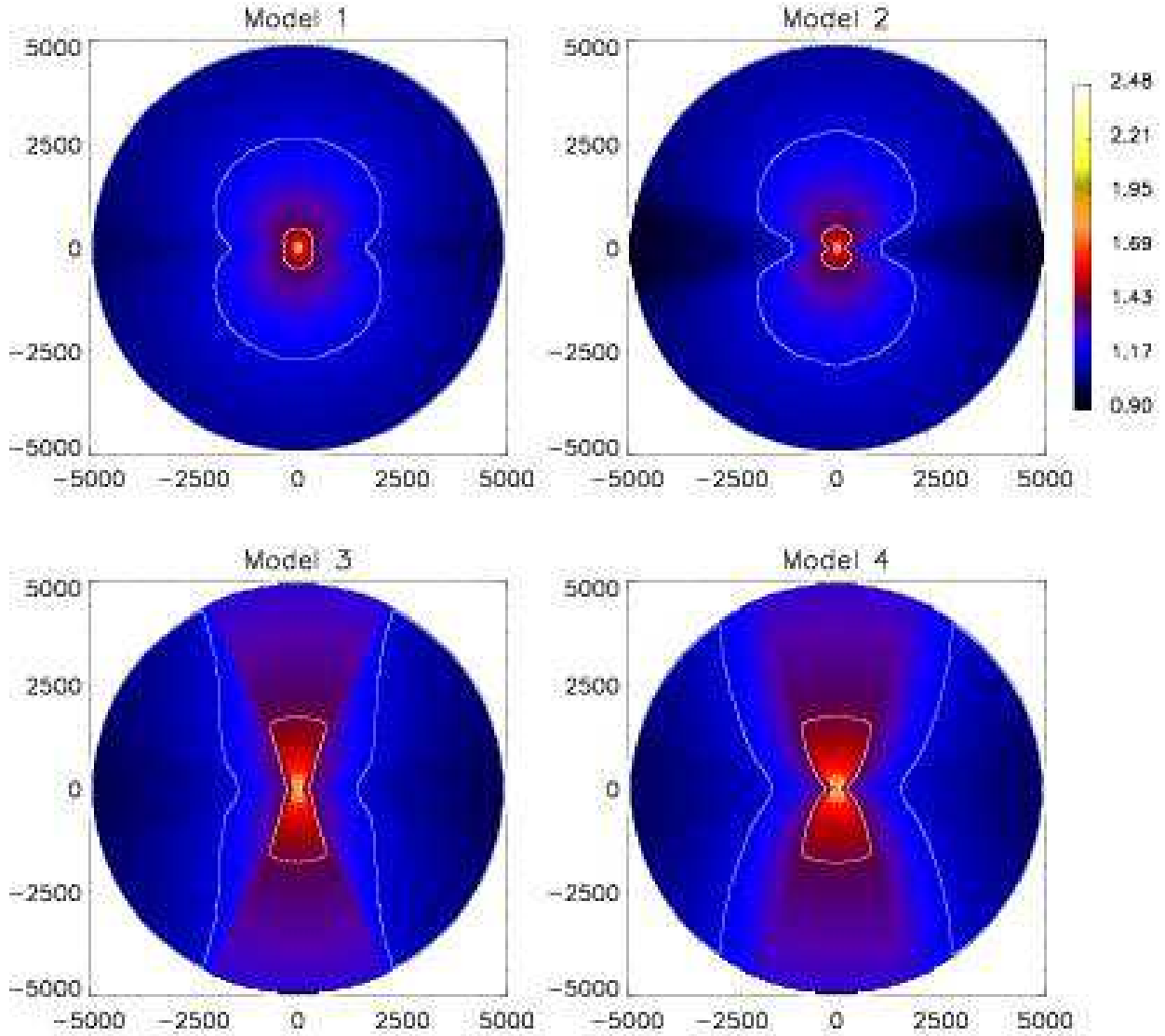}
\caption{Temperature plots for the 4 geometries.  
Images are scaled logarithmically
with the color scale shown at the right.
The contours correspond to the labelled values on the 
color bar (in units of $log(T)$).
The plotted regions and temperature scales are
(a) 5000 AU and $T=8-300$\arcdeg K, 
(b) 300 AU, $T=22-1000$\arcdeg K, and
(c) 20 AU, $T=42-1600$\arcdeg K.
}
\label{t5000}
\end{figure}
\clearpage

\begin{figure}
\figurenum{4b}
\epsscale{1.0}
\plotone{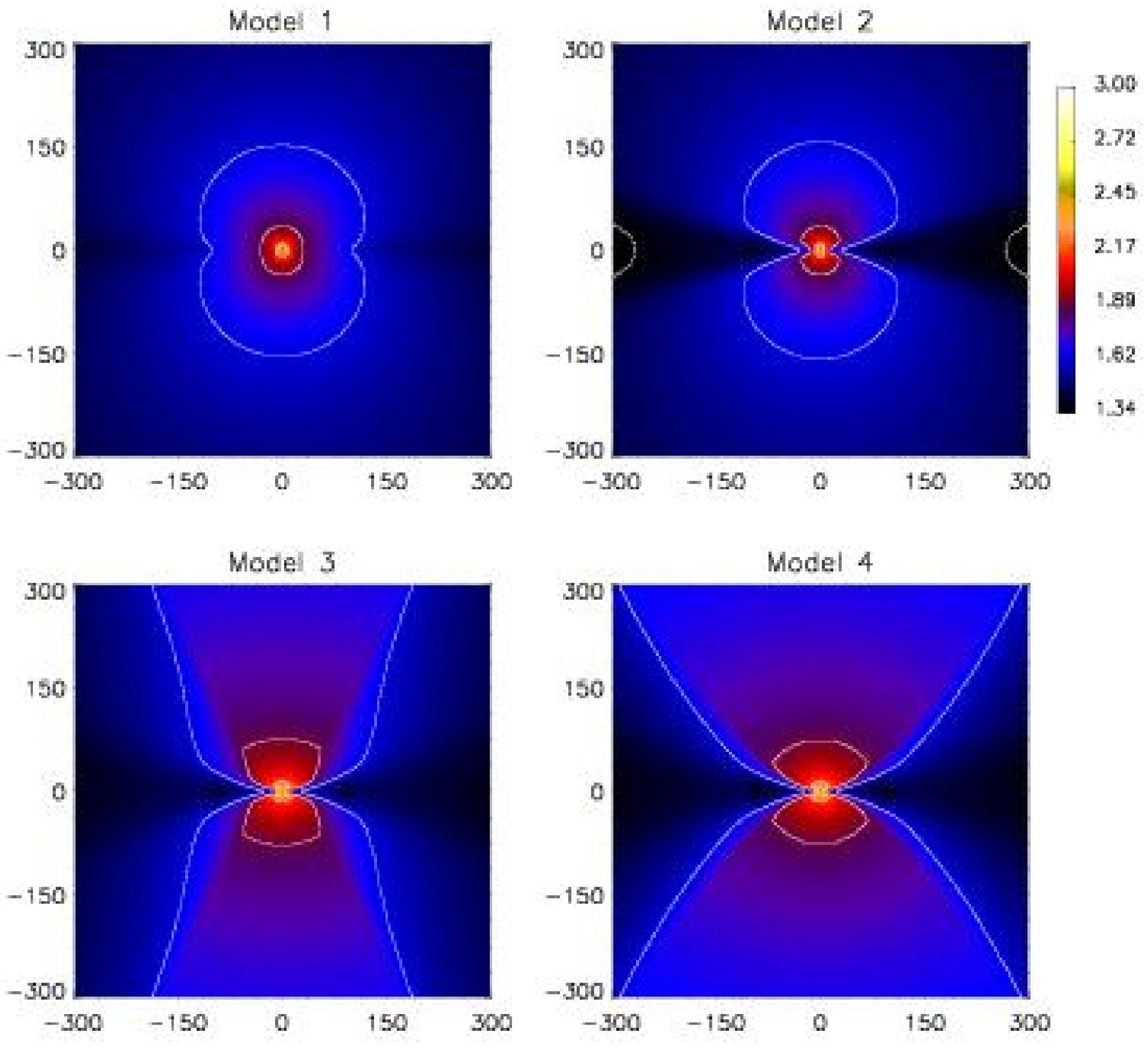}
\label{t300}
\end{figure}
\clearpage

\begin{figure}
\figurenum{4c}
\epsscale{1.0}
\plotone{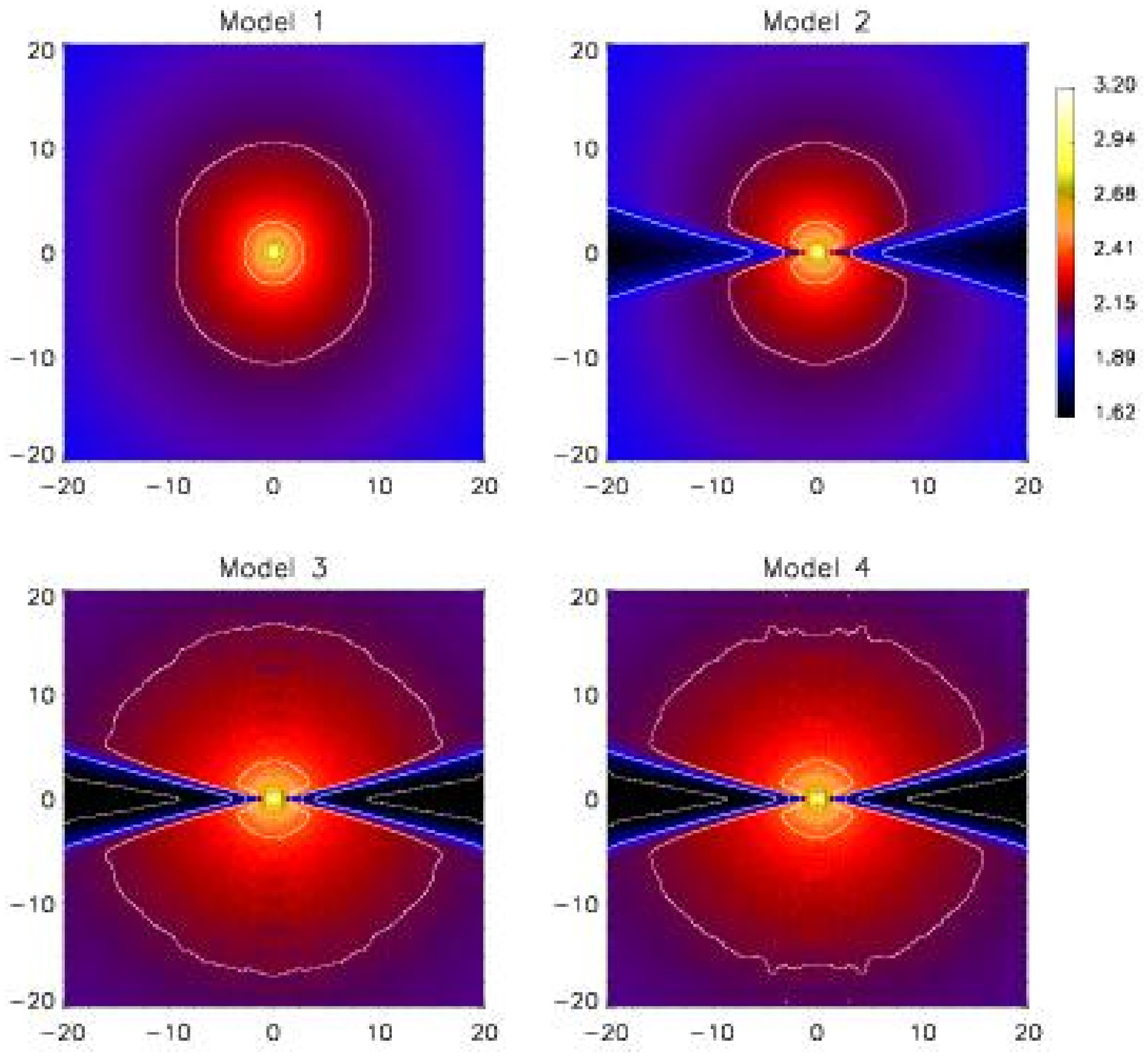}
\label{t20}
\end{figure}
\clearpage

\begin{figure}
\figurenum{5}
\epsscale{1.0}
\plotone{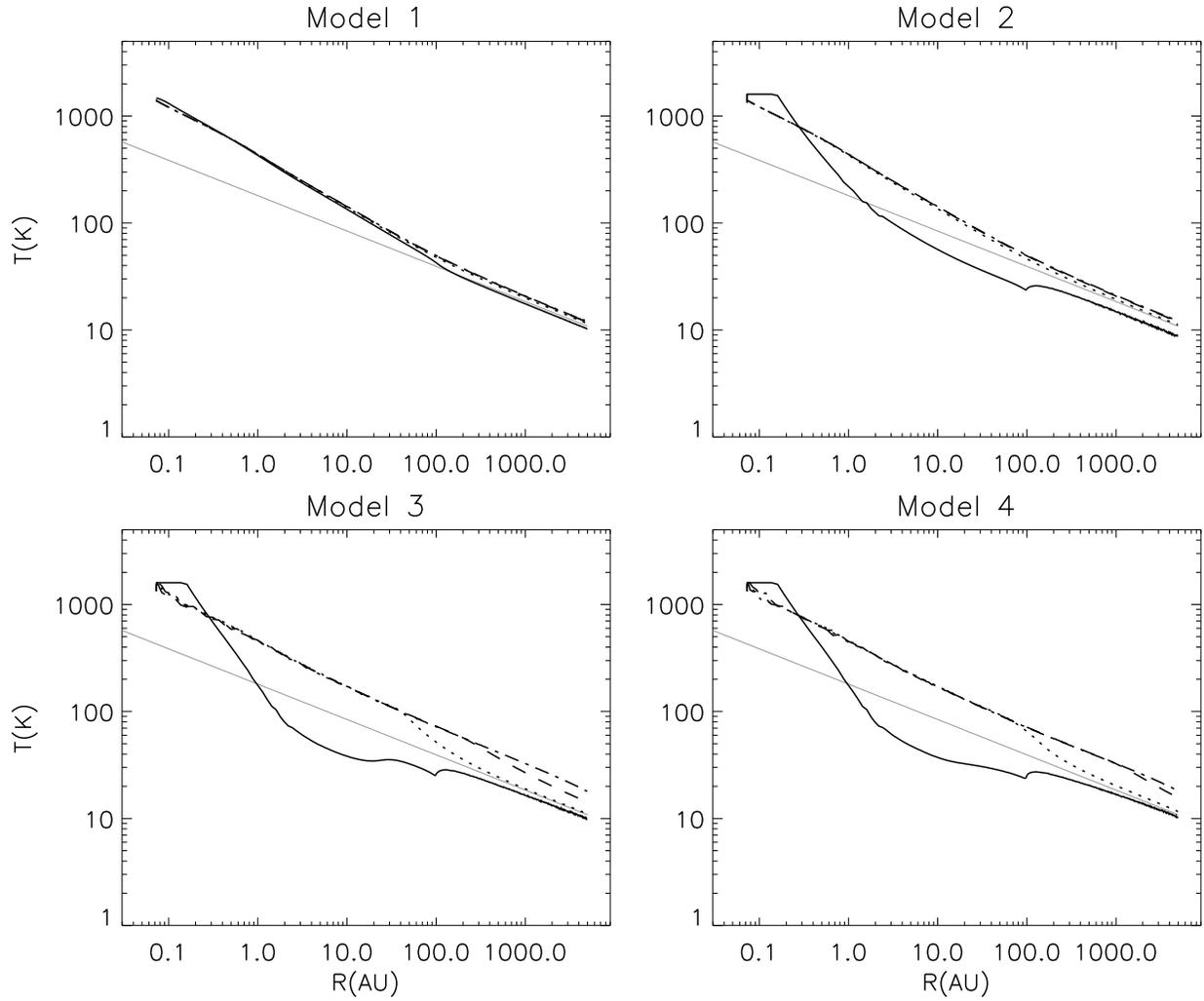}
\caption{
Temperature as a function of radius along four co-latitudes, $\theta$, with
respect to the polar axis:  $\theta=0$ (dot-dashed line), $\theta=30$ (dashed),
$\theta=60$ (dotted), and $\theta=90$ (solid).  The light grey line is a power
law,  
$T = 180 \, (r / 1 {\rm AU})^{-0.33} \, {\rm K}$
}
\label{tr}
\end{figure}
\clearpage

\begin{figure}
\figurenum{6}
\epsscale{1.0}
\plotone{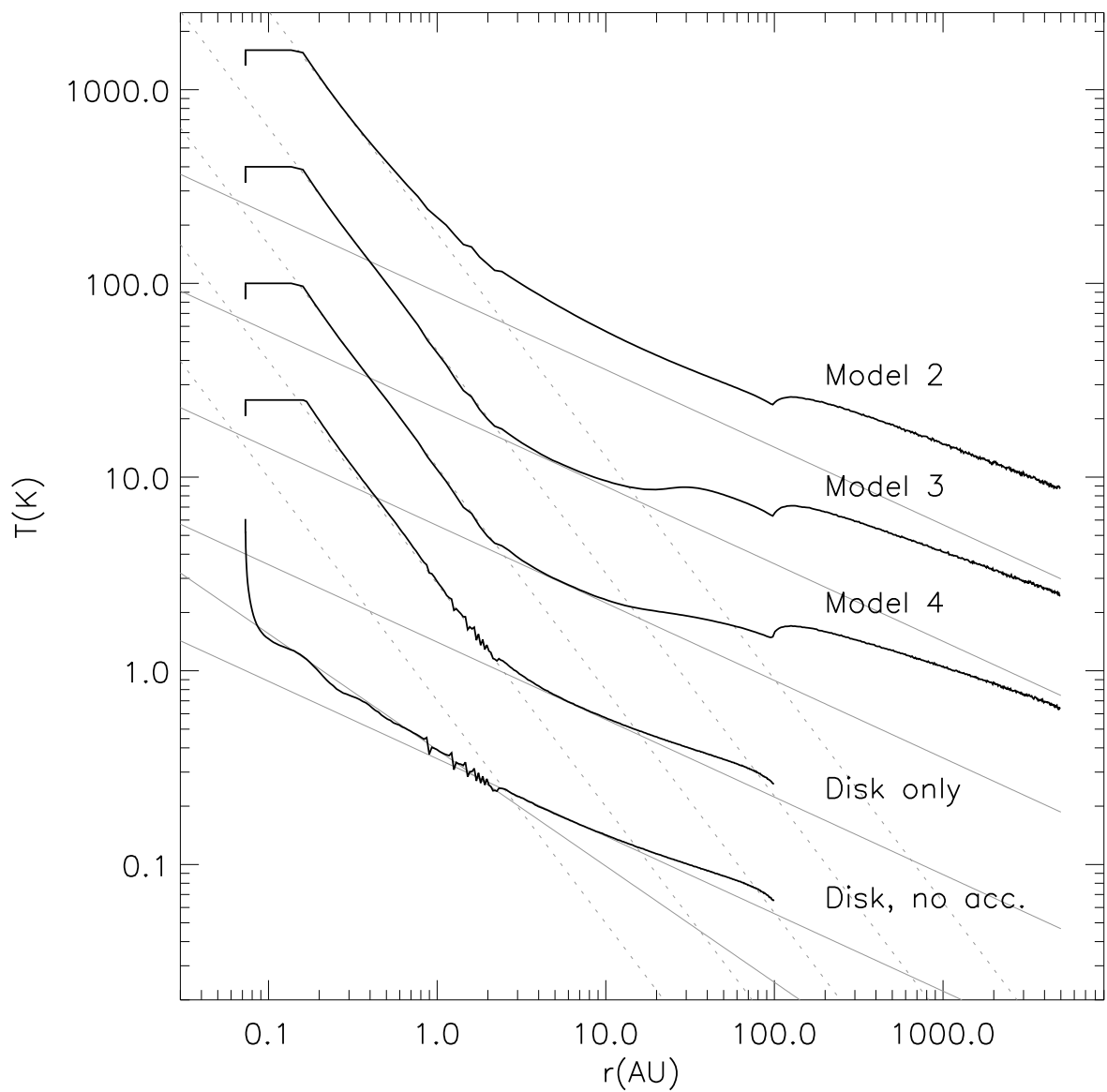}
\caption{
Temperature as a function of radius in the midplane of selected
models with disks.  Each model below the top is scaled down by a factor
of four from the one above.   The grey line is a power law,
$T = 90 \, (r / 1 {\rm AU})^{-0.4} \, {\rm K}$,
also scaled down by a factor of four for each
model.  Notice that the disk midplane is warmer throughout the Model 2 
disk due to
radiation from the surrounding envelope. 
}
\label{tr}
\end{figure}
\clearpage

\begin{figure}
\figurenum{7}
\epsscale{.9}
\plotone{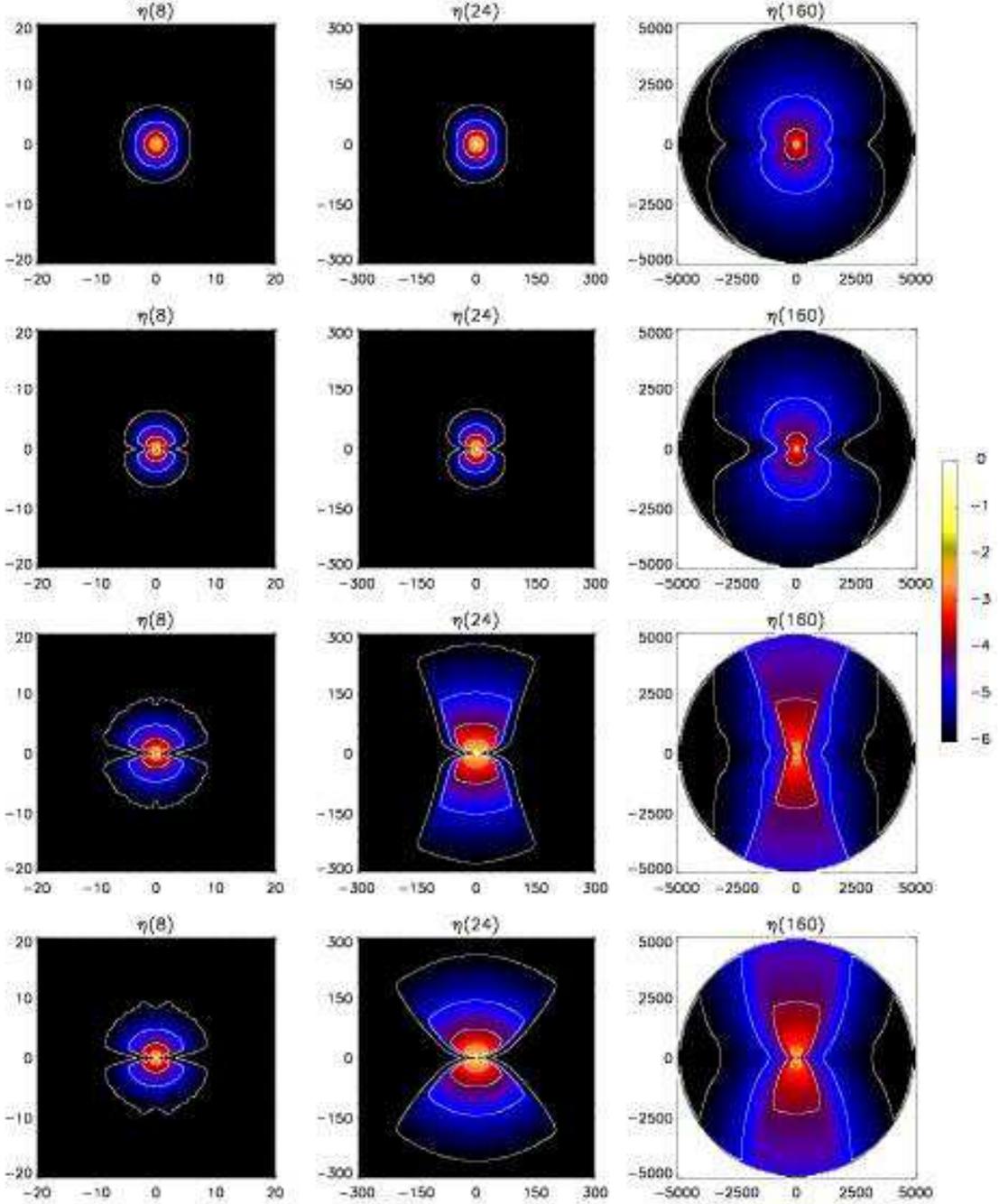}
\caption{Emissivity plots at three wavelengths: 8 $\mu$m (left),
24 $\mu$m (middle) and 
70 $\mu$m (right).  The four model geometries are plotted from top to bottom.
The scaling is logarithmic with the color bar shown
on the right, scaled to the peak.  Peak emissivities plotted are 
$1.5 \times 10^{-1}$ at 8 $\mu$m,
$1.5 \times 10^{-4}$ at 24 $\mu$m, and
$3 \times 10^{-9}$ at 160 $\mu$m,
in units of  erg/cm$^3$/s/Hz/sr.  From left to right the region
plotted spans a radius of 20, 300, and 5000 AU. The emission is more centrally 
concentrated at shorter wavelengths and in models without 
cavities (Models 1, 2).
The contours each decrease by 1
order of magnitude, starting from the peak.}
\label{emiss}
\end{figure}
\clearpage

\begin{figure}
\figurenum{8}
\epsscale{.8}
\plotone{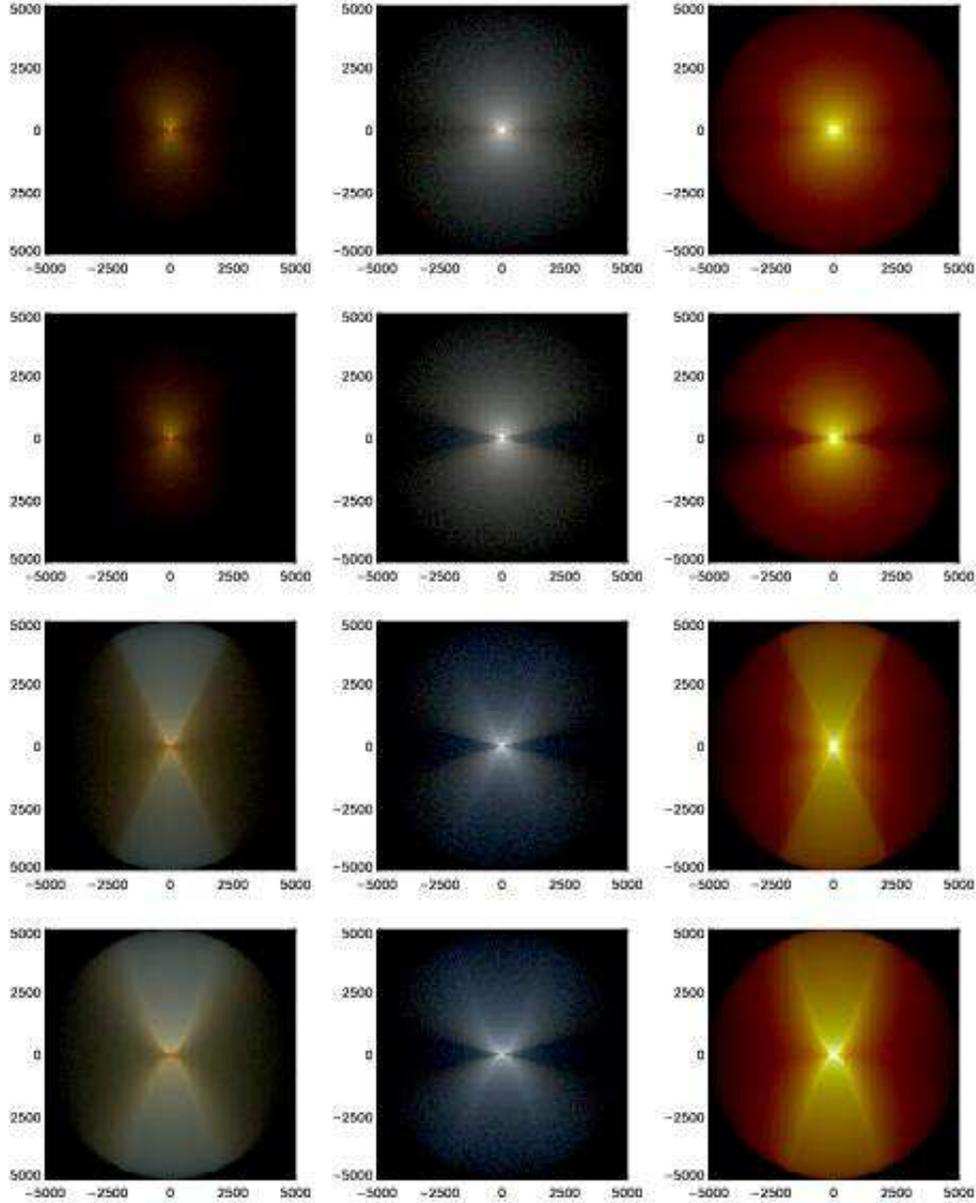}
\caption{
Three-color composite images of the four models (going from top to bottom)
in three different wavelength regions (left to right), viewed at an inclination $i=80\arcdeg$.  
The near-IR images
(left) are composites of images calculated through three NICMOS 
filters centered at
1.1, 1.6 and 2.05 $\mu$m.  The mid-IR images (middle) are composites at three
SIRTF IRAC filters centered at 3.6, 4.5, and 5.8 $\mu$m.  And
the far-IR images (right) are for the SIRTF MIPS filters
centered at 24, 70 and 160 $\mu$m.
The images are scaled logarithmically with minimum intensities of
0.01, 0.0074, and 0.5 MJy/sr for the near-IR, mid-IR, and far-IR images,
respectively, assuming a distance of 140 pc. 
}
\label{imcol}
\end{figure}
\clearpage

\begin{figure}
\figurenum{9}
\epsscale{1.0}
\plotone{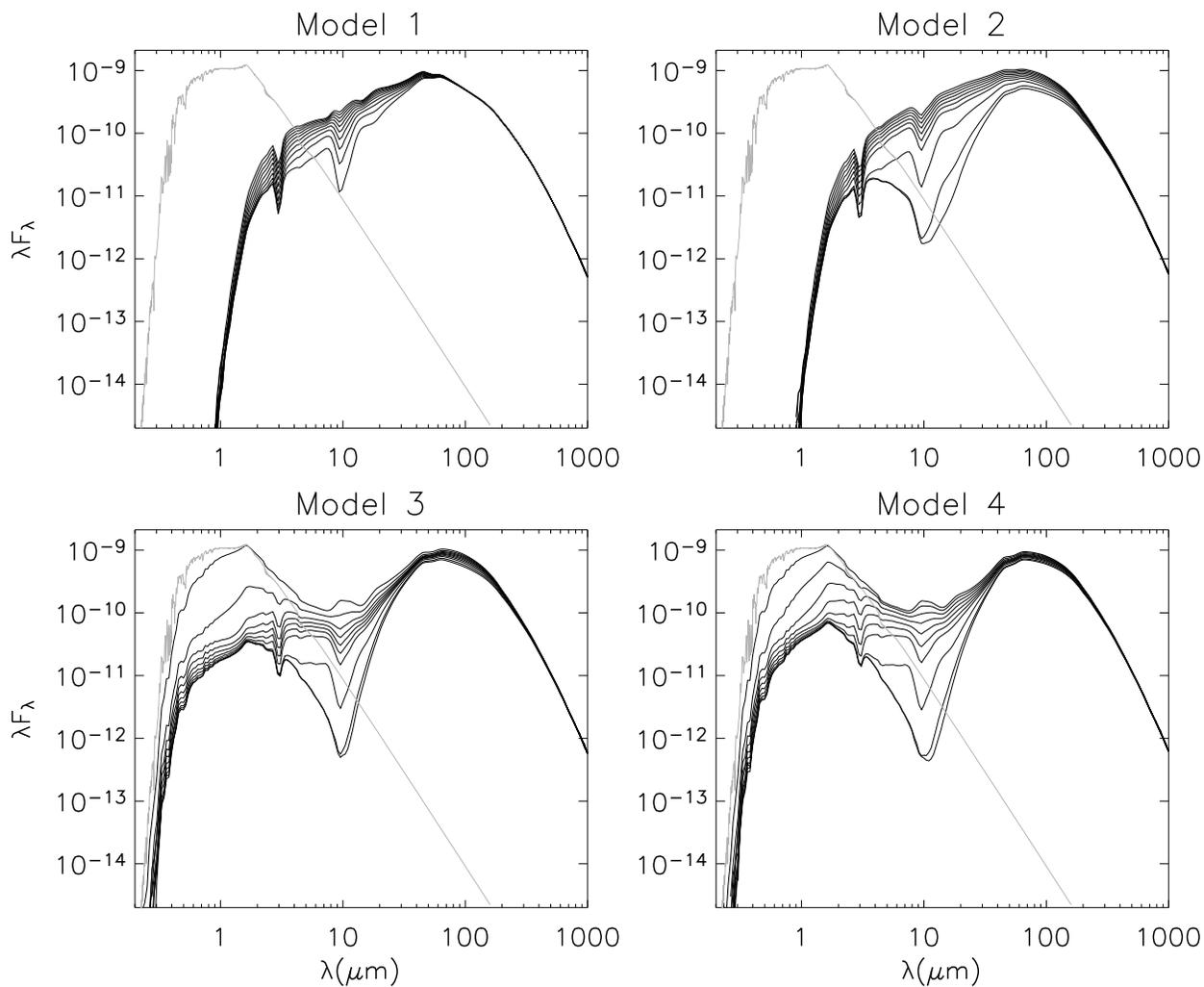}
\caption{
Spectral energy distributions for the four geometries.  Fluxes 
(erg cm$^{-2}$ s$^{-1}$) are 
scaled to a 1 $L_\sun$ source
at a distance of 140 pc.  Ten inclinations are plotted ranging
from $\cos i = 0.05$ to 0.95 at intervals of 0.1.
The light grey line is the input stellar spectrum.  
}
\label{sed4}
\end{figure}
\clearpage

\begin{figure}
\figurenum{10}
\epsscale{1.0}
\plotone{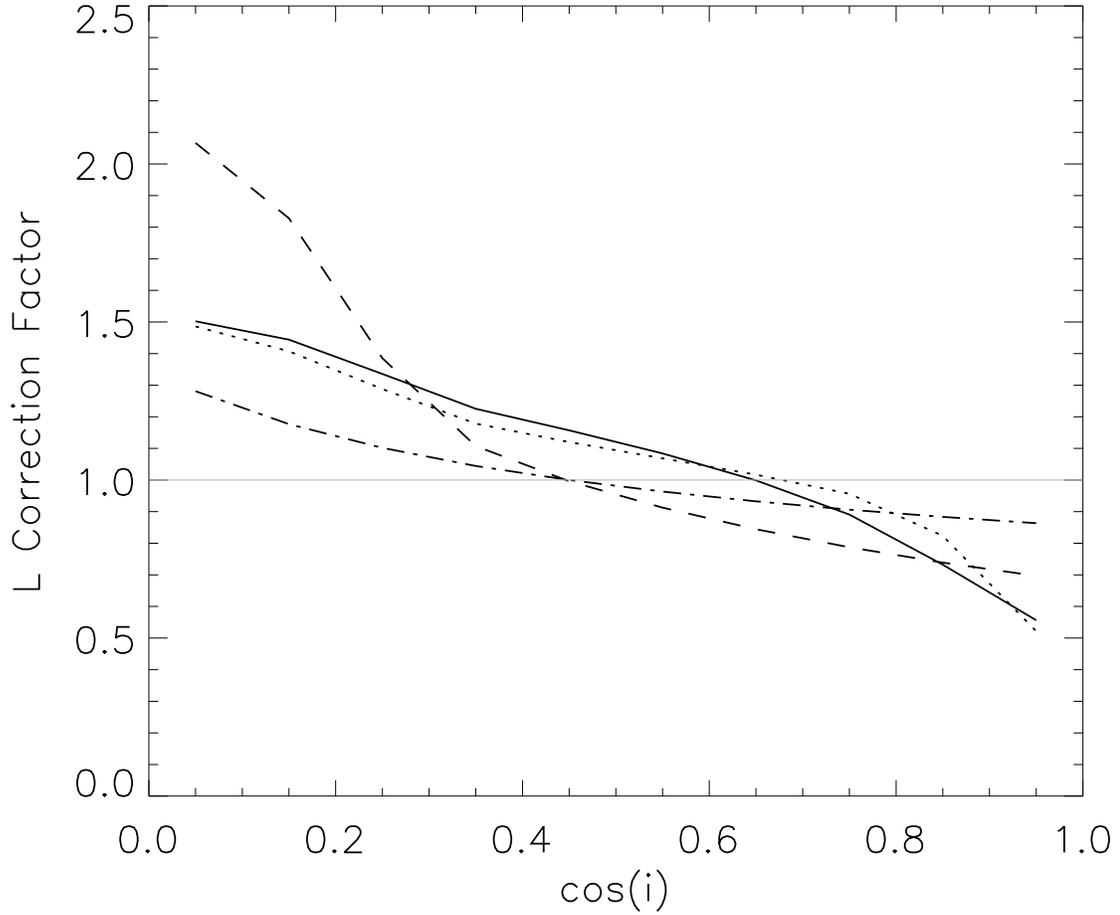}
\caption{
Luminosity correction factors as a function of inclination for the 
four geometries:  Model 1: dot-dashed; Model 2: dashed; Model 3: dotted;
Model 4: solid. 
These curves show the correction factor that would have to be applied to
the integrated SED at a given inclination to correctly determine 
the source luminosity.}
\label{lumi}
\end{figure}
\clearpage

\begin{figure}
\figurenum{11}
\epsscale{1.0}
\plotone{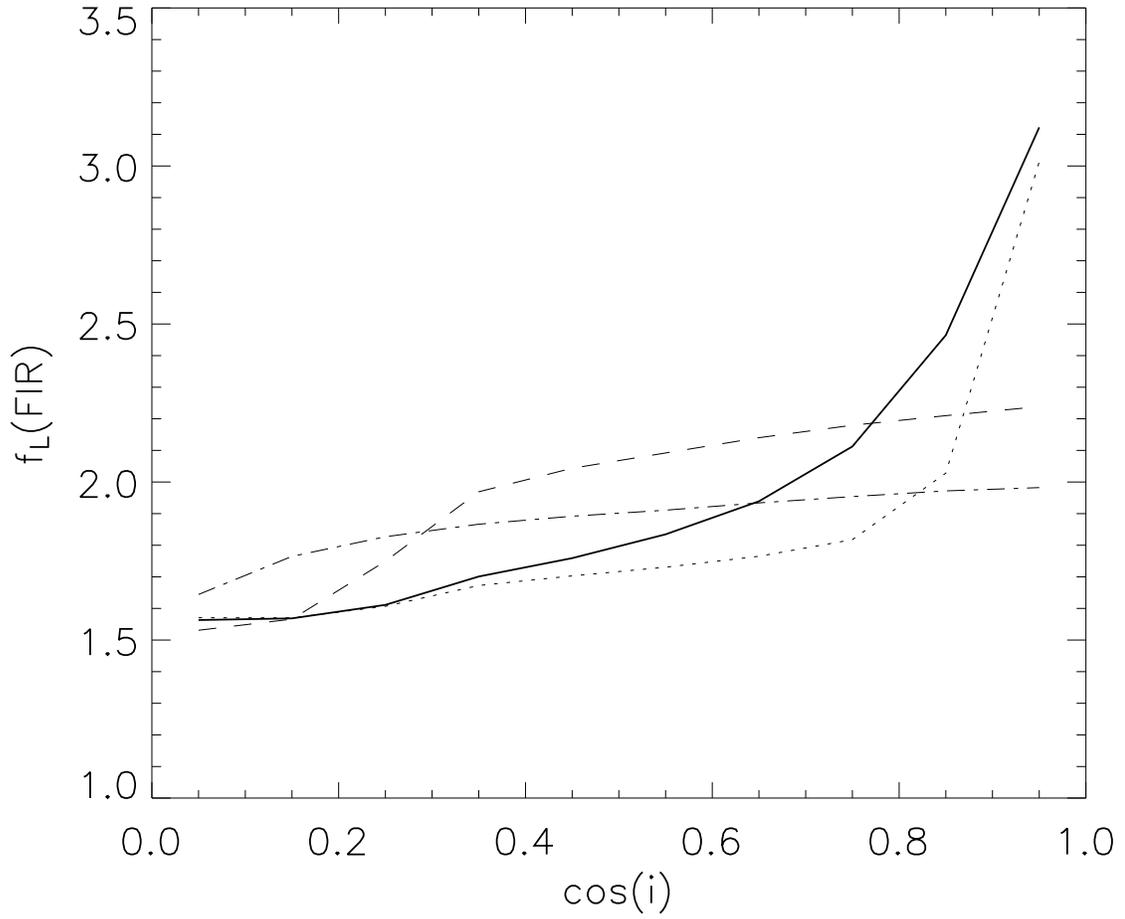}
\caption{
The ratio of the integrated SED to the maximum far-IR flux is plotted
against inclination. Model 1: dot-dashed; Model 2: dashed; Model 3: dotted;
Model 4: solid.}
\label{fl}
\end{figure}
\clearpage

\begin{figure}
\figurenum{12}
\epsscale{1.0}
\plotone{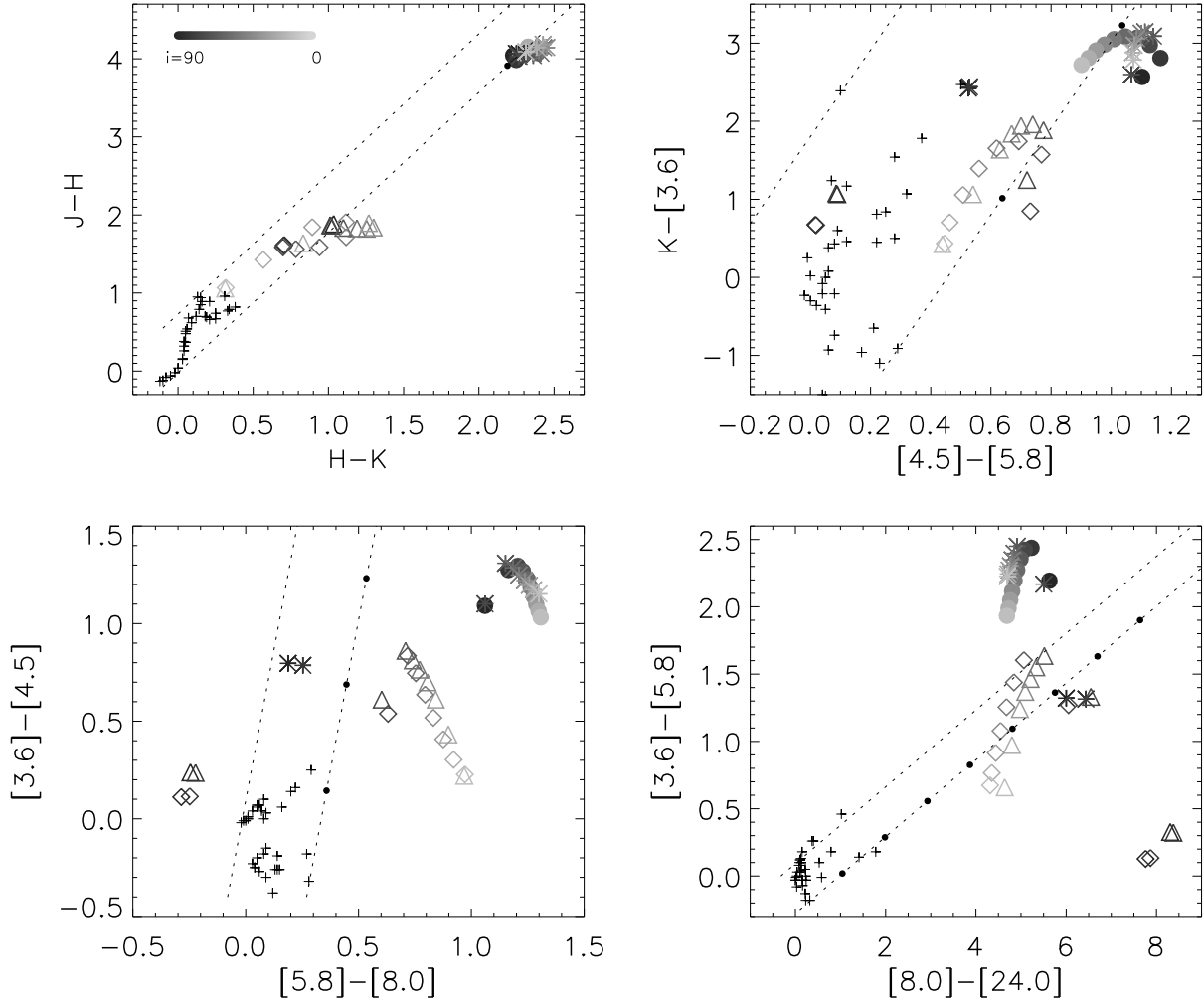}
\caption{
Color-color plots.  
The symbols correspond to the four models as follows, 
Model 1: filled circles; Model 2: asterisks; Model 3: open triangles;
Model 4: open diamonds.
The models with cavities (3 \& 4) are much bluer than
the models without (1 \& 2), despite their similar envelope masses (Table 2).
The symbols are plotted at different levels of greyscale to indicate
inclination, as shown by the greyscale bar in the top left panel.
Edge-on sources are black, and pole-on are light grey.
Notice that some of the bluest sources are edge-on.  This is
due to the dominant contribution of scattered light to the
source flux.  These edge-on sources are very faint so would stand out
in a color-magnitude diagram as faint and blue.
The dotted lines are reddening vectors off the main
sequence and red giant branch (small crosses).  
The black dots along one of the reddening
lines denote increments of A$_V = 30$.  The models in the bottom left
panel are well-separated from reddened main sequence and
giant stars.  The models at bottom right are well-separated from 
stars with A$_V$ less than 100.
}
\label{cc}
\end{figure}
\clearpage

\begin{figure}
\figurenum{13}
\epsscale{1.0}
\plotone{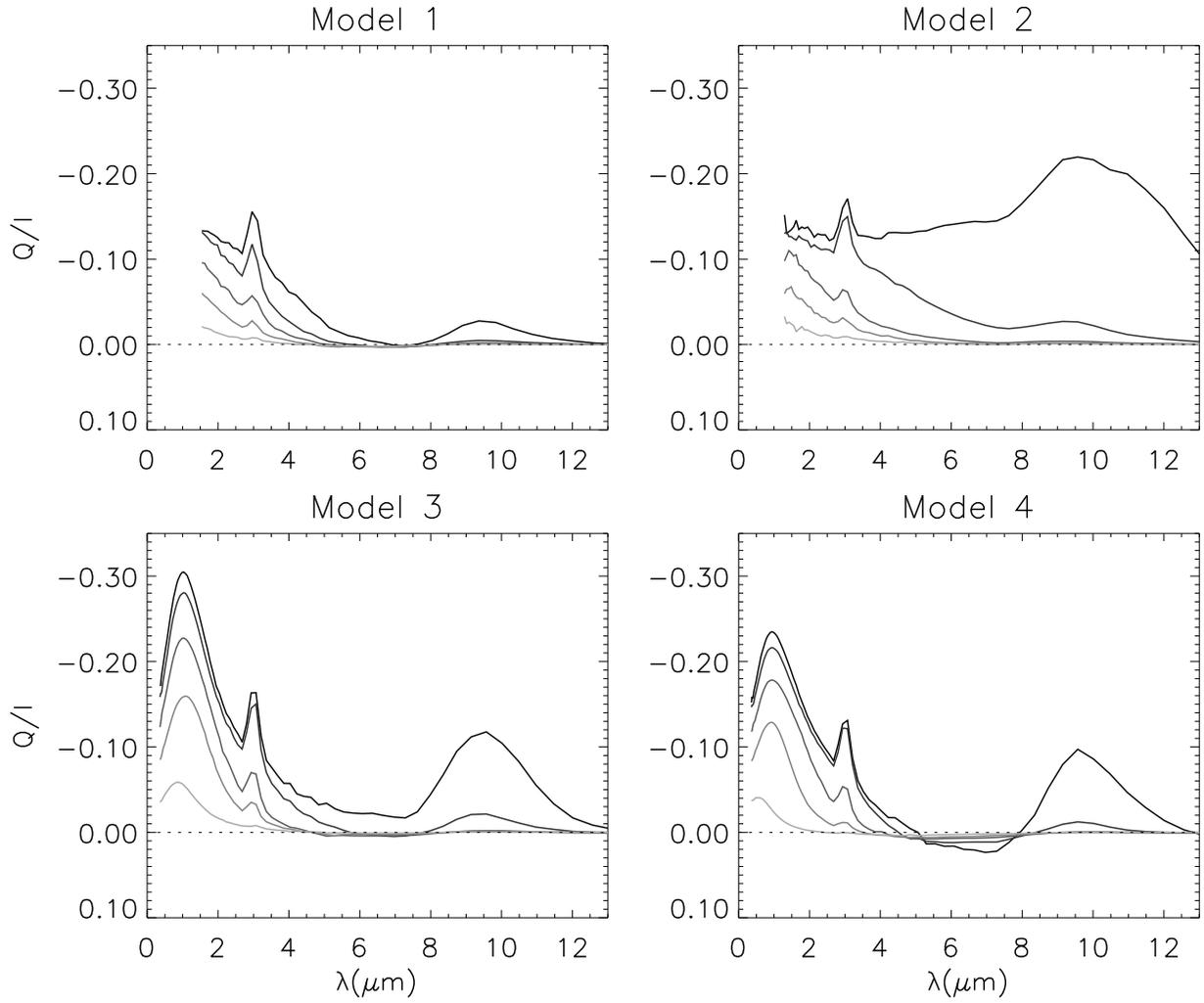}
\caption{
Polarization spectra for the four geometries.  Five inclinations are
plotted from $\cos i=0.05$ to 0.85 in intervals of 0.2.  The lines are
lighter at lower inclinations.
}
\label{sed4}
\end{figure}
\clearpage

\end{document}